\def\breakon{\end{multicols}\widetext\vspace{-.25cm}
\noindent\rule{.48\linewidth}{.3mm}\rule{.3mm}{.5cm}\vspace{0.0cm}}
\def\breakoff{\vspace{-0.5cm}
\noindent
\rule{.52\linewidth}{.0mm}\rule[-.47cm]{.3mm}{.5cm}\rule{.48\linewidth}{.3mm}
\vspace{-0.25cm}
\begin{multicols}{2}   }
\documentstyle[aps,epsf,prb,multicol]{revtex}
\begin{document}

\makeatletter
\renewenvironment{table}
  {\let\@capwidth\linewidth\def\@captype{table}}
  {}

\renewenvironment{figure}
  {\let\@capwidth\linewidth\def\@captype{figure}}
  {}
\makeatother

\title{Opening of the Haldane Gap in Anisotropic Two- and Four-Leg 
Spin Ladders}
\author{Eugene H. Kim }
\address{Department of Physics, University of California,  
         Santa Barbara, California 93106-9530}
\author{J. S\'{o}lyom}
\address{Research Institute for Solid State Physics, P.O. Box 49,
         H-1525 Budapest, Hungary}
\maketitle

\begin{abstract}
We study the opening of the Haldane gap in two-leg and four-leg
anisotropic spin ladders using bosonization and renormalization
group methods, and we determine the phase diagram as a function 
of the interchain coupling and relative anisotropy, $J^z / J^{xy}$. 
It is found that the opening of the Haldane gap is qualitatively 
different for the two cases considered.  For the two-leg ladder 
the Haldane gap opens for arbitrarily small interchain coupling, 
independent of $J^z / J^{xy}$, and the Haldane phase exists in a 
large region of parameter space.  For the four-leg ladder the 
opening of the Haldane gap is strongly dependent on both the 
interchain coupling as well as $J^z / J^{xy}$, and the Haldane 
phase exists only in a narrow region about the isotropic 
antiferromagnet.
\end{abstract}

\vspace{.15in}
\begin{multicols}{2}

\section{Introduction}

Recently the properties of systems which can be modeled by spin ladders
have been extensively studied both experimentally and theoretically
\cite{dagotto}.  It is well established that the magnetic properties of 
these materials depend strongly on the number of legs in the ladder. For
even-leg ladders the susceptibility vanishes exponentially at low 
temperatures, while it shows power-law behavior for odd-leg ladders. This 
difference is due to the fact that the spectrum of magnetic excitations 
is gapless for odd-leg ladders and gapped for even-leg ladders.

This resembles the alternance of a gapless and gapped 
spectrum for isotropic antiferromagnetic spin-$S$ chains, where according 
to Haldane's conjecture\cite{haldane}  the spectrum is gapless if $S$ is 
a half-odd-integer while the spectrum is gapped for integer $S$.  
The two problems are not unrelated \cite{schulz,legeza}, since a 
spin-$S$ chain can be described as $2S$ coupled spin-1/2 chains provided 
the interchain coupling is appropriately chosen. 

Although the appearance of the Haldane gap in an\-iso\-tropic spin-$S$ chains
has been studied by several groups\cite{botet,timonen,gomez,nomura,sakai}, 
most earlier works on coupled chain or ladder models
\cite{hida,scalap,barnes,noack,gopalan} considered only the case 
where the coupling between the spins is isotropic. Some exceptions are the 
work by Watanabe {\sl et al.}\cite{watanabe} using bosonization techniques, 
and by Legeza and S\'olyom \cite{legeza97} where the density matrix 
renormalization group \cite{white} was used to determine the phase diagram 
of anisotropic two-leg ladders. Since the results of the numerical 
calculations were not reliable enough to get a definitive answer, in this 
work we use analytic methods to study the opening of the gap in a ladder 
model when two or four anisotropic spin-1/2 chains are coupled with 
anisotropic interchain couplings.  
Our model is such that when the interchain coupling is equal
to the coupling along the chains, the two-leg (four-leg) ladder behaves
as a spin-1 (spin-2) chain.
    
The rest of the paper is organized as follows.  In Sec.~II the Hamiltonian
of the spin ladder model is presented, and its relationship to the composite
spin representation of spin-$S$ chains is discussed. The model is bosonized 
in Sec.~III, and the results known for spin-1/2 chains are recapitulated.  
The two- and four-leg ladders are analyzed in Secs.~IV and V, respectively, 
and the phase diagram as a function of anisotropy and interchain coupling 
is determined. Finally, in Sec.~VI we summarize and present a discussion 
of our results. The paper also contains two Appendices.  The first 
describes our bosonization conventions, while the second gives a 
derivation of the renormalization group equations.


\section{Coupled Spin Chain Models}

Consider a system of $p$ ($p=2$ or $4$) coupled anisotropic spin-1/2 chains.
Denote the spin operator on chain $\lambda$ ($\lambda = 1,...p$) at site $j$ 
by ${\bf s}_{j,\lambda}$. The spins along the chains are coupled by a 
nearest neighbor Heisenberg exchange; the Hamiltonian for chain $\lambda$ is
\begin{eqnarray}
    H_{\lambda} & = & J^{xy} \sum_j
           \left( s^x_{j,\lambda} s^x_{j+1, \lambda} + s^y_{j, \lambda} 
             s^y_{j+1, \lambda} \right) \nonumber \\
         & & +  J^z \sum_j s^z_{j, \lambda} s^z_{j+1, \lambda} \,.
\label{eq:spin-1/2}
\end{eqnarray}

\begin{figure}
\epsfxsize=3.25in
\centerline{\epsfbox{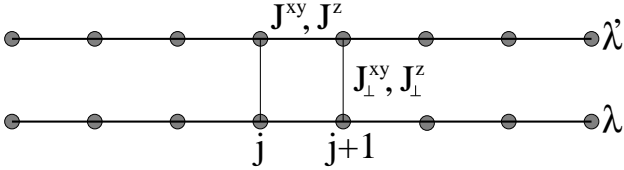}}
\caption{Intra- and interchain couplings in a two-leg ladder}
\label{fig:ladder}
\end{figure}

The coupling between chains can be taken in various forms.  
In spin ladder systems it is most natural to assume that the 
dominant interchain coupling acts between spins on the same rung and 
therefore has the form
\begin{eqnarray}
     H_{\lambda, \lambda'} & = & J_{\perp}^{xy} \sum_j
       \left( s^x_{j, \lambda} s^x_{j, \lambda'} + s^y_{j, \lambda} 
       s^y_{j, \lambda'} \right)    \nonumber \\
      & & + J_{\perp}^z \sum_j s^z_{j, \lambda} s^z_{j, \lambda'} \,.
\label{eq:H-lambda}
\end{eqnarray}
This kind of the interchain coupling is shown schematically in Fig. 
\ref{fig:ladder}.

When the interchain coupling is strong enough and ferromagnetic, the
ground state and the low lying excited states of the ladder are those
in which the $p$ spins on rung $j$ add up to form a larger spin of length 
$p/2$. Thus in

\vspace{.2in}
\noindent
this limit the two-leg ladder is equivalent to a spin-1 chain in some 
of its physical properties, while the four-leg ladder gives an 
effective spin-2 chain. Strictly speaking this equivalence is valid only 
in the limit when the rung coupling becomes very strong.  Nevertheless, it 
was established that in the isotropic two-leg ladder the Haldane gap is 
generated for arbitrarily small interchain coupling \cite{watanabe}, i.e., 
its critical value is zero. 

\begin{figure}
\epsfxsize=3.25in
\centerline{\epsfbox{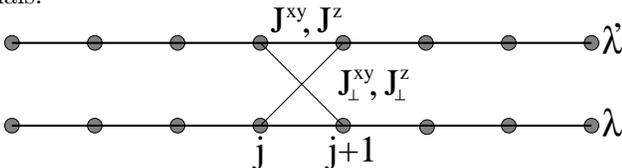}}
\caption{Intra- and interchain couplings in a composite-spin model}
\label{fig:cross}
\end{figure}

Alternatively, one could use the composite spin representation for
spin chains\cite{timonen}. This representation is obtained by taking a single 
spin-$S$ chain with Hamiltonian
\begin{equation}
 H  = J^{xy} \sum_j
           \left( S^x_{j} S^x_{j+1} + S^y_{j} S^y_{j+1} \right)
           +  J^z \sum_j S^z_j S^z_{j+1}
\end{equation}
and writing the spin operator as a sum of $p=2S$ spin-1/2 operators
\begin{equation}
 S^{\alpha}_j  = s_{j,1}^{\alpha} + s_{j,2}^{\alpha}
                     + \dots + s_{j,p}^{\alpha} \,.
\label{eq:composite}
\end{equation}
Inserting this into the equation above, we get
\begin{equation}
 H  =  \sum_{\lambda=1}^p H_{\lambda} + \sum_{\lambda, \lambda' \atop
   \lambda < \lambda'} H_{\lambda  \lambda'} \,,
\label{eq:ham-comp}
\end{equation}
where $H_{\lambda}$ is the same as Eq. (\ref{eq:H-lambda}), so the first
part is the Hamiltonian of $p$ spin-1/2 Heisenberg chains, while the 
interchain coupling between them has the form
\begin{eqnarray}
 H_{\lambda \lambda'} & = &  J_{\perp}^{xy} \sum_j
        \left( s^x_{j,\lambda} s^x_{j+1,\lambda'} +
        s^x_{j,\lambda'} s^x_{j+1,\lambda}  \right. \nonumber \\
      & & \phantom{++++} + \left.  s^y_{j,\lambda} s^y_{j+1,\lambda'} +
         s^y_{j,\lambda'} s^y_{j+1,\lambda} \right)  \nonumber \\
        &  & +  J_{\perp}^{z} \sum_j \left( s^z_{j,\lambda} s^z_{j+1,\lambda'} 
         + s^z_{j,\lambda'} s^z_{j+1, \lambda} \right) \,,
\label{eq:h-lam-lam'}
\end{eqnarray}
with
\begin{equation}
      J_{\perp}^{xy} = J^{xy} \,, \quad J_{\perp}^{z} = J^{z} \,.
\end{equation}
One can view this, as shown schematically in Fig. \ref{fig:cross}, as a 
ladder in which the interchain coupling is along diagonals.

In this composite spin model the Hilbert space is larger than in the 
original spin-$S$ model. For example, when two spin-1/2's are added, 
in addition to the symmetric combinations that give the triplet $S=1$ state, 
the antisymmetric combination can also appear that corresponds to a singlet, 
$S=0$ state. It has been shown \cite{timonen}, however, that the ground state 
and the low lying excited levels of this composite-spin model are identical to 
that of the true $S=1$ model. The configurations where an $S=0$ appears on 
at least one site have much higher energies than those we are interested in, 
since they effectively break the chains.  Although it has not been checked 
numerically, it is reasonable that the same holds for the four-chain case,
i.e., the lowest lying levels are those where the model is equivalent to 
the spin-2 chain; configurations where $S=0$ or $S=1$ appear on at least
one site effectively break the homogeneous chain, and therefore have 
higher energies.

In actual ladder materials, in contrast to the cases discussed above, 
the interchain coupling seems to be antiferromagnetic and act between 
nearest neighbor spins on the same rung.  However, White showed that 
the two-leg ladder with (isotropic) antiferromagnetic coupling between 
nearest neighbor spins is in the same universality class as the spin-1 
chain \cite{white96}.
 
\begin{figure}
\epsfxsize=3.1in
\centerline{\epsfbox{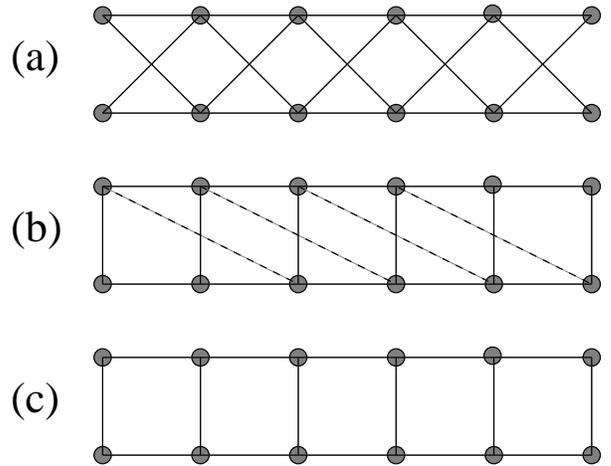}}
\caption{White's argument showing that the (isotropic) antiferromagnetic
two-leg ladder and the spin-1 chain are in the same universality class.
(a) Start with the composite spin representation for a spin-1 chain.  
(b) Slide ``chain-2'' by one lattice spacing to the left, and gradually 
turn off the coupling between sites $\sqrt{5}$ away.  (c) We are left with 
an antiferromagnetic two-leg ladder.}
\label{fig:white2leg}
\end{figure}

White's arguments were the following.  Start with the composite spin 
representation for the spin-1 chain.  Take ``chain 2'' and slide 
it by one lattice spacing to the left.  Now, gradually turn off the 
coupling between sites $\sqrt{5}$ away; denote this coupling by 
$J_{\perp}'$.  White was able to show that the system with 
$J_{\perp}' = 0$ evolved continuously from the system with 
$J_{\perp}' = J_{\perp}$.  (i.e., There was no change in symmetry 
and no disappearance or appearance of gaps \cite{white96}.)  
(See Fig.~\ref{fig:white2leg}.)  

White's arguments can be applied to the four-leg ladder.  
For now, let us only consider isotropic couplings.  (The case
of anisotropic couplings will be discussed later.)

\vspace{.125in}

\vspace{.2in}
\noindent  
Start with the composite spin repersentation for the 
spin-2 chain.  Shift ``chain 2'' and ``chain 4'' to the left by one 
lattice spacing.  Then, gradually turn off the interchain couplings
between sites which are not nearest neighbors.  We are left with
an antiferromagnetic ladder with only nearest neighbor couplings.
Therefore, as long as we have no change in symmetry and no 
disappearance or appearance of gaps as we turn off the couplings, 
the antiferromagnetic four-leg ladder will be in the same universality 
class as the spin-2 chain.  It seems reasonable that the two are related 
by continuity.  Numerical studies have shown the antiferromagnetic 
four-leg ladder to have a gapped spectrum \cite{morewhite,poilblanc}.  
It is hard to believe that, while performing the transformation above, 
we could go from one gapped phase to another {\it independent} gapped 
phase, and cross a critical point in the process.  Therefore, it is 
reasonable that the four-leg (isotropic) antiferromagnetic ladder and 
the spin-2 chain are in the same universality class.  

\begin{figure}
\epsfxsize=3.25in
\centerline{\epsfbox{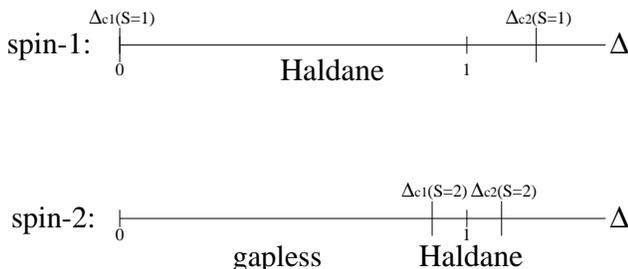}}
\caption{Lower and upper critical anisotropies for the Haldane 
phase for $S=1$ and $S=2$.}
\label{fig:crit-val}
\end{figure}

One could ask how the Haldane phase of the ladder arises when the 
interchain coupling is switched on between anisotropic spin-1/2 chains
which are gapless for $-1 \leq \Delta \leq 1$ and become antiferromagnetic
with finite gap for $\Delta > 1$. It is known that for anisotropic chains
with integer spin, the Haldane phase shrinks to narrower and narrower ranges 
around the isotropic antiferromagnetic point as the spin length increases. 
Let $\Delta = J^z/J^{xy}$. Denote the lower boundary of the Haldane phase 
for spin-S by $\Delta_{c1}(S)$ and the upper boundary by $\Delta_{c2}(S)$.
Then
\begin{equation}
     \Delta_{c1}(S_1) < \Delta_{c1}(S_2) < 1 < \Delta_{c2}(S_2) < 
       \Delta_{c2}(S_1) \,,
\end{equation}
for $S_1 < S_2$. Numerical calculations give $\Delta_{c1}(S=1) = \varepsilon$ 
($\varepsilon > 0$, $\varepsilon \ll 1$) and $\Delta_{c1}(S=2) \approx 0.9$. 
In the continuum model of the spin-1 chain, $\varepsilon =0$ is obtained
\cite{dennijs}. The phase boundaries of the Haldane phase are shown 
schematically for the spin-1 and spin-2 chains in Fig. \ref{fig:crit-val}. 
Therefore, the interesting range of anisotropy for the ladder is 
$0 < \Delta \leq 1$, where the spin-1/2 chain is gapless, while the 
spin-1 and spin-2 chains are in the Haldane phase for some range of $\Delta$.  

Starting from the gapless situation and coupling the chains, taking into 
account that at the isotropic antiferromagnetic point the critical value 
of the interchain coupling is zero\cite{barnes}, there are two natural ways

\vspace{.15in}

\begin{figure}
\centerline{\epsfxsize=1.7in \epsfbox{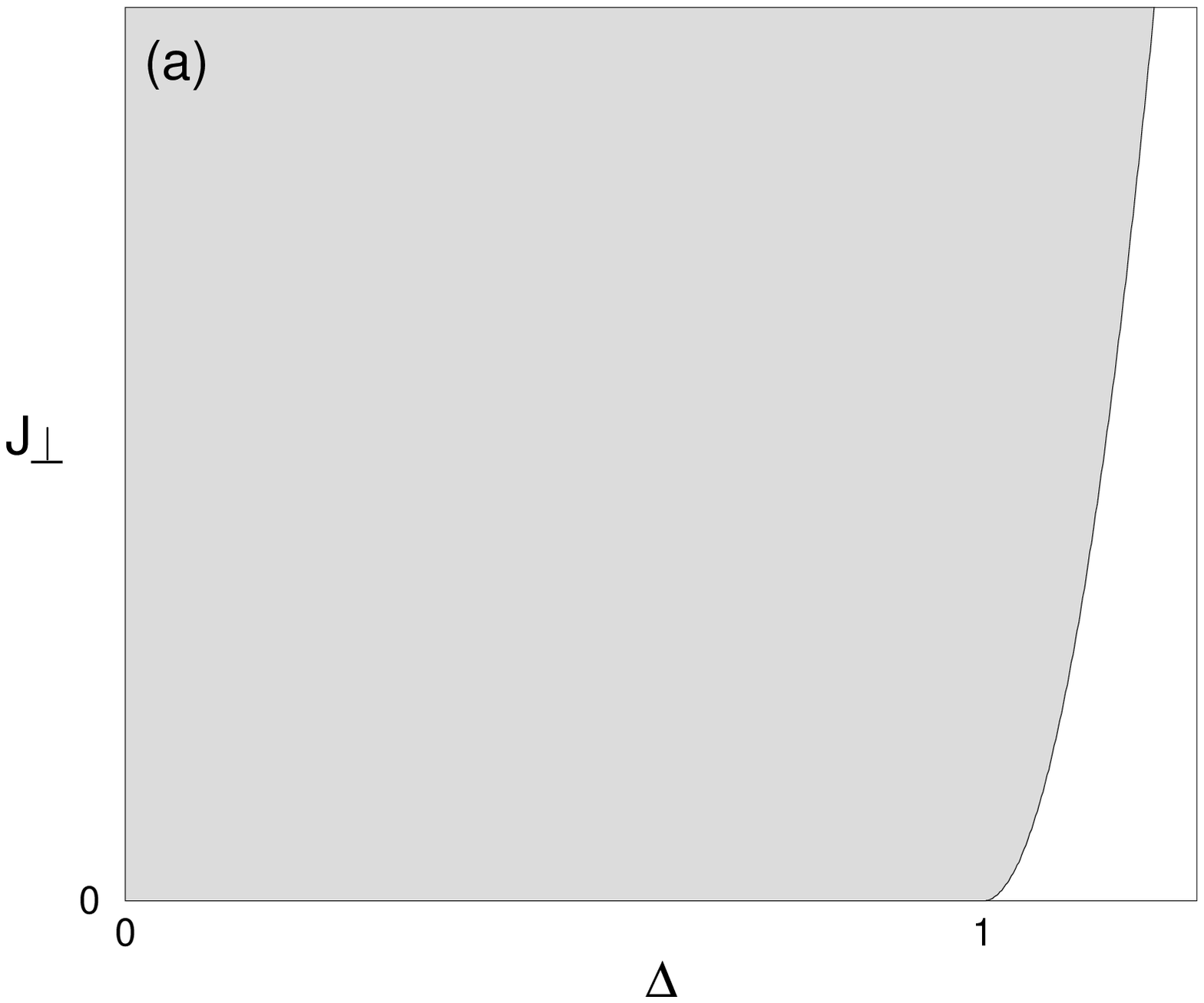}
            \epsfxsize=1.7in \epsfbox{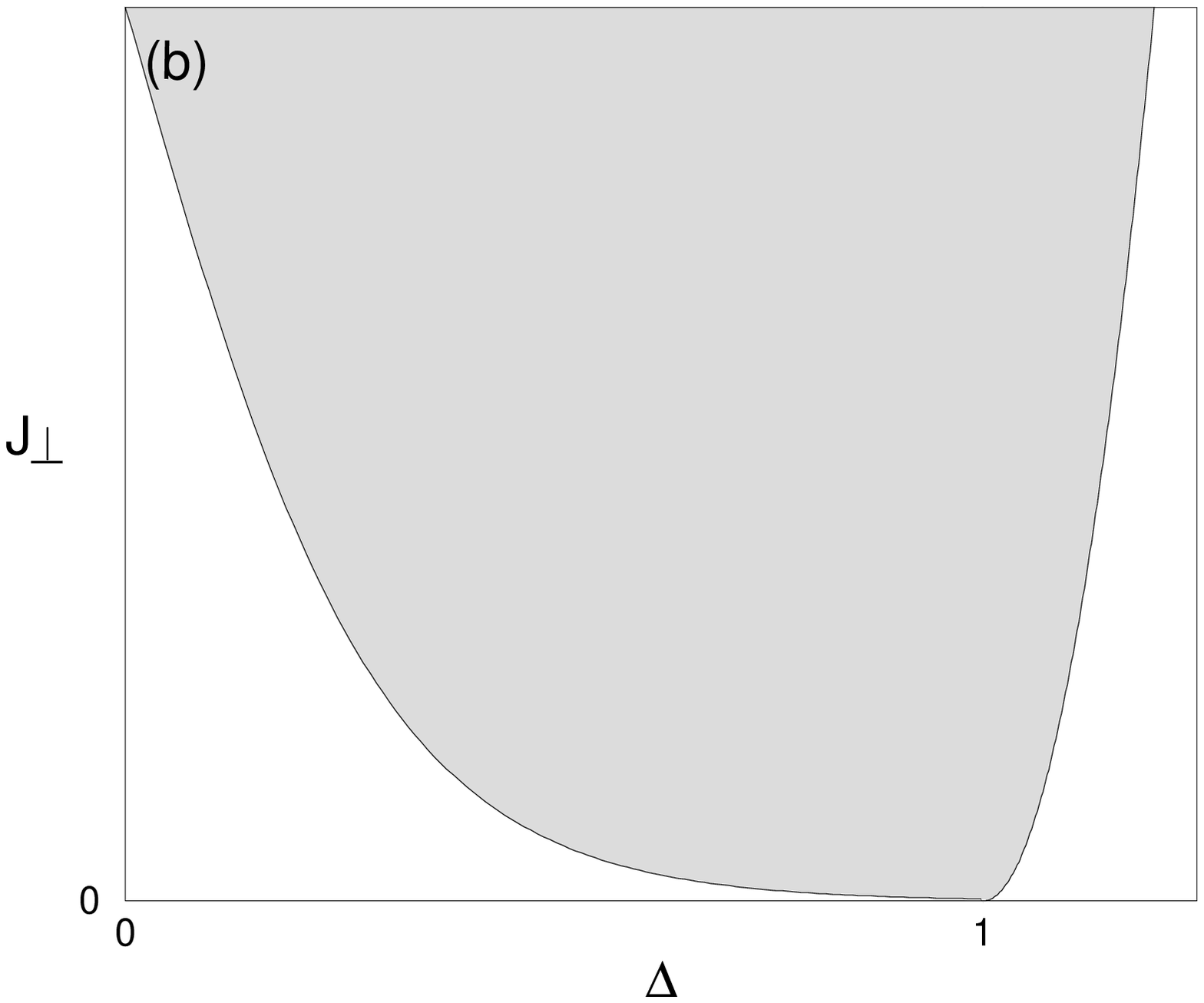}}
\centerline{\epsfxsize=1.7in \epsfbox{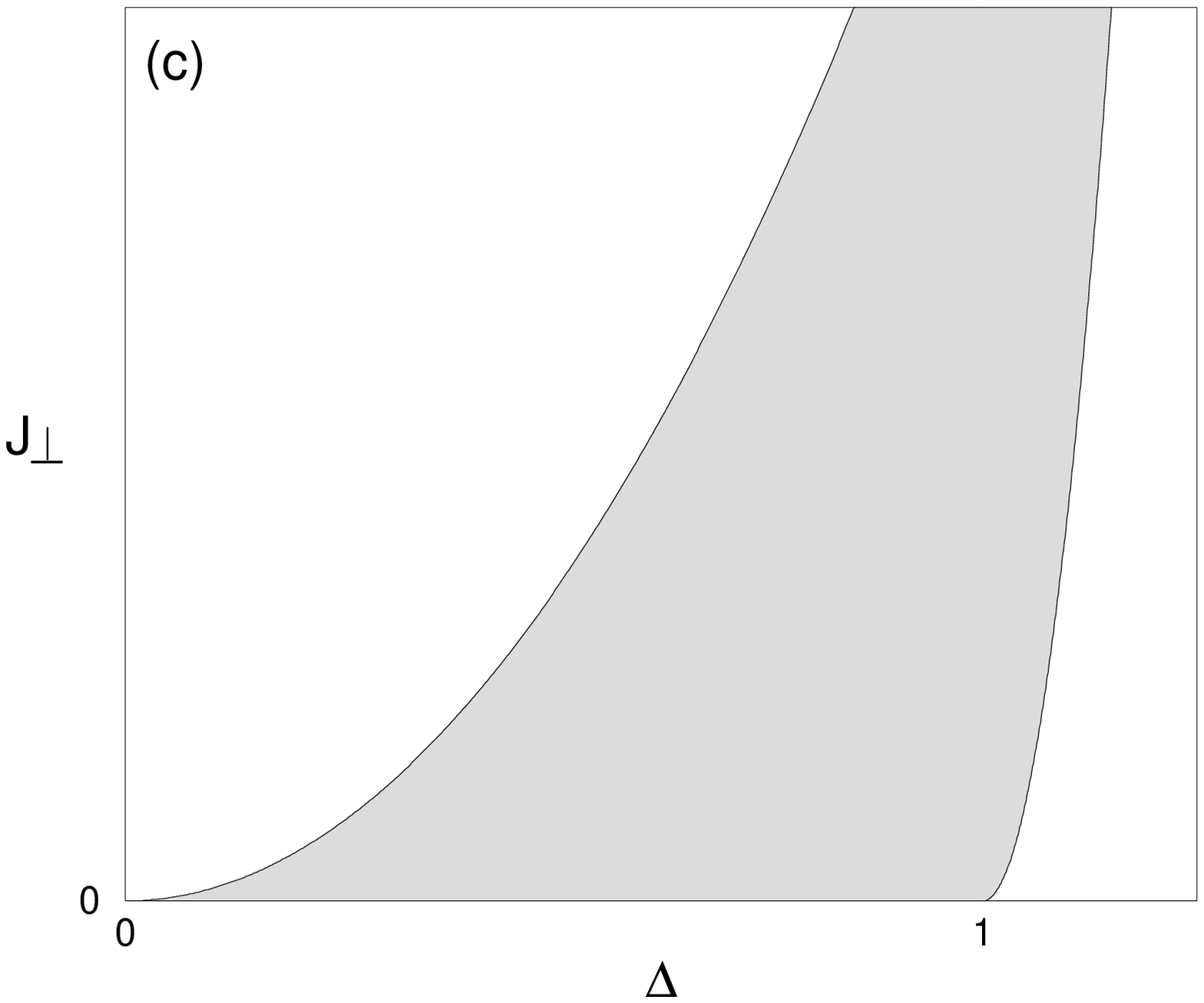}
            \epsfxsize=1.7in \epsfbox{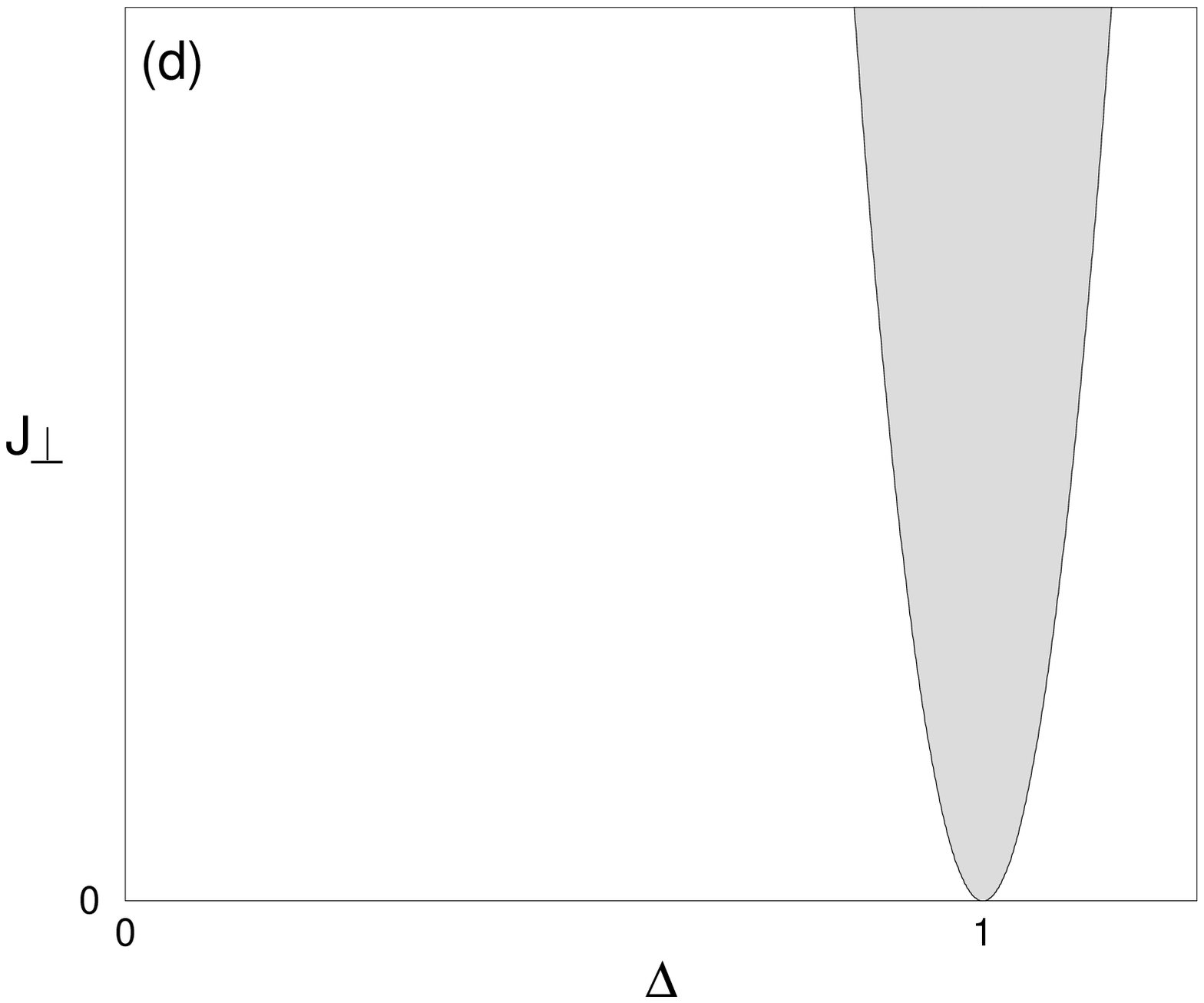}}
\caption{Possible scenarios for the opening of the Haldane gap.  (a) \& (b)
Scenarios for the two-leg ladder.  (c) \& (d) Scenarios for the four-leg 
ladder.  The shaded region is the Haldane phase.}
\label{fig:expect}
\end{figure}

\vspace{.15in}
\noindent
for the opening of the Haldane gap as the interchain 
coupling is switched on, as shown in Fig. \ref{fig:expect}.
Our aim is to decide which 
scenario occurs.  For computational reasons, for the rest of the paper
we will work with the composite spin model.  In order to study 
how our system evolves from two (four) decoupled spin-1/2
chains into a ladder that is equivalent to a spin-1 (spin-2) chain, 
as well the roles of the planar 
and Ising like parts of the interchain coupling individually,
the Hamiltonian in Eq. (\ref{eq:ham-comp}) will be generalized by 
allowing for $J_{\perp}^{xy} \neq J^{xy}$ and $J_{\perp}^z \neq J^{z}$ in Eq. 
(\ref{eq:h-lam-lam'}).


\section{Bosonizing Spin Chains}

While the spin-1/2 Heisenberg chain can be solved exactly by the Bethe-Ansatz,
models with higher spin values can be treated, even in one-dimension, 
only approximately. One such approximation is to express the spin model 
in terms of fermions or bosons and develop a scaling theory for these 
degrees of freedom. In this section, we outline the key 
ingredients needed to fermionize and bosonize spin models. 

\subsection{The Spin-1/2 Chain}

First consider a single anisotropic spin-1/2 chain described by the 
Hamiltonian (\ref{eq:spin-1/2}), where the chain index $\lambda$ will be
dropped. The Jordan-Wigner transformation 
\begin{eqnarray}
  s^{+}_j = s_j^x + i s_j^y & = & c^{\dagger}_j  
    \exp \left( i \pi \sum_{l=0}^{j-1} c^{\dagger}_l c^{\phantom \dagger}_l 
       \right)  \,,  \nonumber \\
   s^{-}_j  = s_j^x - i s_j^y & = & \exp \left( - i \pi \sum_{l=0}^{j-1} 
    c^{\dagger}_l c^{\phantom \dagger}_l \right) c^{\phantom \dagger}_j \,, \\
  s^z_j & = & c^{\dagger}_j c^{\phantom \dagger}_j - {\textstyle \frac{1}{2}} 
      \,,    \nonumber     
\end{eqnarray}
or because of the fermionic nature of the $c^{\phantom \dagger}_j, 
c^{\dagger}_j$ operators the equivalent form
\begin{equation}
  s^{+}_j = c^{\dagger}_j \cos \left( \pi \sum_{l=0}^{j-1} c^{\dagger}_l 
        c^{\phantom \dagger}_l  \right)  ,  \quad
   s^{-}_j  = \cos \left( \pi \sum_{l=0}^{j-1} c^{\dagger}_l 
     c^{\phantom \dagger}_l \right) c^{\phantom \dagger}_j , 
\label{affleckwigner}
\end{equation}
allows us to write this Hamiltonian in terms of spinless fermions 
\begin{eqnarray}
 H & = & {\textstyle \frac{1}{2}} J^{xy} \sum_j  \left( c^{\dagger}_{j} 
    c^{\phantom \dagger}_{j+1} +  c^{\dagger}_{j+1} c^{\phantom \dagger}_{j} 
     \right)  \nonumber \\
    & &  +  J^z  \sum_j \left( c^{\dagger}_{j} c^{\phantom \dagger}_{j} 
     - {\textstyle \frac{1}{2}} \right) \left( c^{\dagger}_{j+1} 
      c^{\phantom \dagger}_{j+1} - {\textstyle \frac{1}{2}} \right)  \,.
\end{eqnarray}
After Fourier transforming to momentum space
\begin{eqnarray}
    H & = & J^{xy} \sum_k \cos (ka) c^{\dagger}_{k} 
       c^{\phantom \dagger}_{k} 
      \nonumber \\
      & &  + { J^z \over N}  \sum_q \cos (qa) \rho(q) \rho(-q) \,,
\end{eqnarray}
where $a$ is the lattice constant, $N$ the number of sites in the chain, and
\begin{equation}
     \rho(q) = \sum_k c^{\dagger}_{k+q} c^{\phantom \dagger}_k  \,.
\label{density}
\end{equation}
In the spin liquid phase where there is no net magnetic moment, the band is 
half filled. 

To study the low-energy properties of the model, we can
linearize the spectrum around the Fermi points, $k_{F} = \pm \pi/2a$.
In this way a Luttinger model is obtained, where $v_{F} = J^{xy} a$ is the 
unrenormalized velocity. In what follows we will work with a rescaled 
Hamiltonian, dividing $H$ by $J^{xy} a$, and use a dimensionless coupling 
$\Delta = J^{z}/ J^{xy}$ (the relative anisotropy).  

Since the low-energy excitations of the Luttinger liquid
are density fluctuations (which are bosons), it is convenient to use, in the
continuum limit, a further transformation from fermions to bosons.
For clarity and completeness, we have given a description of our 
bosonization conventions in Appendix A.

For $-1 \leq \Delta \leq 1$ where the spin-1/2 Heisenberg chain
is known to be critical, the continuum bosonized form of the Hamiltonian
is
\begin{equation}
    H = \frac{u}{2} \int dx \left[ K \Pi^2  + 
         \frac{1}{K} (\partial_x \Phi)^2 \right] \, .
\label{free-boson}
\end{equation}
For $\mid \Delta \mid \ll 1$ the parameters $u$ and $K$ can be determined
perturbatively.  However, the general form of $u$ and $K$ can be obtained
by comparison with the Bethe-Ansatz solution \cite{luther,tsvelik}. 
They are given by
\begin{equation}
 K = \frac{\pi}{2 \left( \pi - \arccos \Delta \right)} , \qquad  
 u = \frac{\pi \sqrt{1 - \Delta^2} } {2 \arccos \Delta} .
\end{equation}
An important property is that for 
 $\Delta = 0$ (i.e., the free fermion point) $K = 1$, and for
 $\Delta = 1$ (i.e., the isotroic point) $K = 1/2$.

We will also need to know the bosonized form for the spin operators.
Using the bosonization rules of Appendix A, in the continuum limit they can 
be shown to be 
\begin{eqnarray}
   s^{+}(x) & = & { s^{+}_j \over \sqrt{a}}  = \frac{ \exp\left( -i \sqrt{\pi} 
     \Theta  \right) } {\sqrt{2\pi a} } \left[ e^{-i(\pi x/a)} + 
      \cos(\sqrt{4\pi} \Phi ) \right] ,   \nonumber \\
   s^z(x) & = & { s^z_j  \over a } = \frac{1}{\sqrt{\pi}} \partial_x \Phi 
       + e^{i (\pi x/a)} \frac{ \sin( \sqrt{4\pi} \Phi) }{\pi a} .
\label{spin-boson}
\end{eqnarray}


\subsection{Coupled Spin-1/2 Chains} 

For coupled spin-1/2 chains the same procedure can be used to express
the spin operators in terms of boson fields, by simply attaching a 
$\lambda$ index to the field. The intrachain part of the Hamiltonian
will have the same form as Eq.~(\ref{free-boson}). 
After rescaling by the bare velocity, $J^{xy}a$, we write the Hamiltonian 
of the interchain coupling term in the form
\begin{eqnarray}
 H_{\lambda \lambda'} & = &  {1 \over a} J_{\perp}^{xy} \sum_j
        \left( s^x_{j,\lambda} s^x_{j+1,\lambda'} +
        s^x_{j,\lambda'} s^x_{j+1,\lambda}  \right. \nonumber \\
      & & \phantom{+++} + \left.  s^y_{j,\lambda} s^y_{j+1,\lambda'} +
         s^y_{j,\lambda'} s^y_{j+1,\lambda} \right)  \nonumber \\
        &  & + {1 \over a} J_{\perp}^{z} \sum_j \left( s^z_{j,\lambda} 
     s^z_{j+1,\lambda'} + s^z_{j,\lambda'} s^z_{j+1, \lambda} \right) \,,
\label{eq:h-lam-lam'-2}
\end{eqnarray}
where we have let
\[
     {J^{xy}_{\perp} \over J^{xy} } \rightarrow  J^{xy}_{\perp} \,
     \quad  {\rm and} \quad   
     {J^{z}_{\perp} \over J^{xy} } \rightarrow  J^{z}_{\perp} \,.
\]
Using the bosonized form for the spin operators, we get 
\begin{eqnarray}
 H_{\lambda,\lambda'} & = & \int {dx \over (2 \pi a)^2} \left[ 
  g_1 \cos \left( \sqrt{4\pi} (\Phi_{\lambda} + \Phi_{\lambda'}) \right) 
     \right. \nonumber \\
  & & \phantom{++} + g_2 \cos \left( \sqrt{4\pi} (\Phi_{\lambda} - 
      \Phi_{\lambda'})  \right) \nonumber \\
  & & \phantom{++} \left. + g_3 \cos \left( \sqrt{\pi} (\Theta_{\lambda} - 
     \Theta_{\lambda'}) \right)   \right]   \nonumber \\
 & & +  \frac{2 J_{\perp}^z }{\pi} \int dx 
 \partial_x \Phi_{\lambda} \partial_x \Phi_{\lambda'}  \\
   & & +  g_4 \int {dx \over (2 \pi a)^2} \cos \left( \sqrt{\pi} 
    (\Theta_{\lambda} -  \Theta_{\lambda'}) \right)  \nonumber \\
  & & \phantom{++} \times \cos \left( \sqrt{4\pi} 
    \left( \Phi_{\lambda} + \Phi_{\lambda'} \right) \right)  \nonumber  \\
  & & +  g_5 \int {dx \over (2 \pi a)^2} \cos \left( \sqrt{\pi} 
    (\Theta_{\lambda} -  \Theta_{\lambda'}) \right) \nonumber \\
    & & \phantom{++} \times \cos \left( \sqrt{4\pi} 
    \left( \Phi_{\lambda} - \Phi_{\lambda'} \right) \right) ,  \nonumber
\end{eqnarray}
where 
\begin{eqnarray}
 g_1 = 4 J_{\perp}^z \,, & \quad &  g_2 = - 4 J_{\perp}^z \,, \nonumber \\
 g_3 = -4 \pi J_{\perp}^{xy} \,, & \quad &  g_4 = g_5 = 2 \pi J_{\perp}^{xy} 
    \,.
\end{eqnarray}

Note that the coupling between the $x$ and $y$ components of the spins 
on different chains is non-local.


\section{ The Two-Leg Ladder }

We first consider how the Haldane gap is generated in the anisotropic
two-leg ladder when two gapless spin-1/2 chains are coupled by an
interchain coupling of the form given in Eq. (\ref{eq:h-lam-lam'}). 
The Hamiltonian of the system is 
\begin{equation}
 H  = H_1  +  H_2  +  H_{1,2} \, .  
\end{equation}
As mentioned before for $J_{\perp}^{xy} = 1 $ and $J_{\perp}^z = J^z$,
our Hamiltonian is equivalent, as far as the low-lying states are concerned,
to the true spin-1 Heisenberg chain. Therefore, we have a model which 
interpolates between two uncoupled spin-1/2 chains and a spin-1 chain.

The bosonized form of the Hamiltonian is 
\begin{eqnarray}
  H  & = &  \frac{u}{2} \int dx 
     \left[ K \Pi_1^2  + \frac{1}{K} (\partial_x \Phi_1)^2 \right] \nonumber \\
    & &  + \frac{u}{2} \int dx  \left[ K \Pi_2^2  + \frac{1}{K} 
      (\partial_x \Phi_2)^2 \right]   \nonumber \\
    & &  +  \frac{2 J_{\perp}^z }{\pi} \int dx 
         \partial_x \Phi_1 \partial_x \Phi_2   \nonumber \\
   & & + \int {dx \over (2 \pi a)^2}  
   \left[ g_1 \cos \left( \sqrt{4\pi} (\Phi_1 + \Phi_2) 
     \right)  \right. \nonumber \\ 
    & & \phantom{++} + g_2  \cos \left( \sqrt{4\pi} (\Phi_1 - \Phi_2) 
       \right) \nonumber \\  
     & & \phantom{++} + \left. g_3  \cos \left( \sqrt{\pi} (\Theta_1 - 
        \Theta_2) \right) \right]  \nonumber \\
  & & + \int \frac{dx}{(2\pi a)^2} 
   \left[ g_4 \cos \left( \sqrt{\pi} (\Theta_1 - \Theta_2) \right) \right.
    \nonumber \\
    & & \phantom{++} \times \cos \left( \sqrt{4\pi} (\Phi_1 + \Phi_2) \right)
    \nonumber \\
   & & \phantom{+} + g_5 \cos \left( \sqrt{\pi} (\Theta_1 - \Theta_2) \right) 
      \nonumber \\
       & & \left. \phantom{++}  \times  \cos \left( \sqrt{4\pi} 
       (\Phi_1 - \Phi_2) \right)  \right] \,.
\end{eqnarray}

It is useful to define the fields 
\begin{equation}
 \Phi_s = \frac{1}{\sqrt{2}} \left( \Phi_1 + \Phi_2 \right) \, ,  \ \ \ 
 \Phi_a = \frac{1}{\sqrt{2}} \left( \Phi_1 - \Phi_2 \right) \,.
\end{equation}
In terms of these fields our Hamiltonian has the form
\begin{eqnarray}
   H  & = &  \frac{u_s}{2} \int dx \left[ K_s \Pi_s^2  + 
         \frac{1}{K_s} (\partial_x \Phi_s)^2 \right] \nonumber \\
      & & +  g_1 \int {dx \over (2 \pi a)^2} 
         \cos \left( \sqrt{8\pi} \Phi_s \right)   \nonumber \\
    & & +  \frac{u_a}{2} \int dx  \left[ K_a \Pi_a^2  +  
         \frac{1}{K_a} (\partial_x \Phi_a)^2 \right]  \nonumber \\
    & & + \int {dx \over (2 \pi a)^2} 
    \left[ g_2 \cos  \left( \sqrt{8\pi} \Phi_a \right) 
     +  g_3 \cos \left( \sqrt{2\pi} \Theta_a \right) \right] \nonumber \\
    & & + g_4 \int \frac{dx}{(2\pi a)^2} 
    \cos \left( \sqrt{2\pi} \Theta_a \right)
    \cos \left( \sqrt{8\pi} \Phi_s \right)   \nonumber \\
    & & + g_5 \int \frac{dx}{(2\pi a)^2} 
    \cos \left( \sqrt{2\pi} \Theta_a \right)
    \cos \left( \sqrt{8\pi} \Phi_a \right)   \,,
\end{eqnarray}
where 
\begin{eqnarray}
 K_s = K \left( 1 + \frac{2KJ_{\perp}^z}{u \pi} \right)^{-1/2} \,,  
  &  & \ \ \ 
 u_s = u \left( 1 + \frac{2K J_{\perp}^z}{u \pi} \right)^{1/2} \,, \nonumber \\
    & & \\
 K_a = K \left( 1 - \frac{2KJ_{\perp}^z}{u \pi} \right)^{-1/2} \, ,
    & & \ \ \ 
 u_a = u \left( 1 - \frac{2K J_{\perp}^z}{u \pi} \right)^{1/2} \, . \nonumber 
\end{eqnarray}
For $J_{\perp}^{xy}$,$J_{\perp}^z \ll 1$ we have
\begin{eqnarray}
 K_s \approx K \left( 1 - \frac{K J_{\perp}^z}{u \pi} \right),
 & &  \ \
 u_s \approx u \left( 1 + \frac{K J_{\perp}^z}{u \pi} \right) , 
 \nonumber \\
 K_a \approx K \left( 1 + \frac{K J_{\perp}^z}{u \pi} \right),
  &  &  \ \
 u_a \approx u \left( 1 - \frac{K J_{\perp}^z}{u \pi} \right) .
\end{eqnarray}

We are interested in whether or not the interchain coupling causes 
a gap in the excitation spectrum.  Therefore, we would like to identify
the relevant operators; these operators will ``pin'' their arguments,
thus causing gaps to appear.  To do this we consider the scaling
dimensions of the operators in the interchain coupling (see Appendix A).  
The scaling dimension of  $\cos \left( \sqrt{8\pi} \Phi_s \right)$ 
is $2K_s$; the scaling dimension of 
$\cos \left( \sqrt{8\pi} \Phi_a \right)$ is $2K_a$; 
the scaling dimension of $\cos \left( \sqrt{2\pi} \Theta_a \right)$ is 
$1/(2K_a)$; the scaling dimension of  
$\cos \left( \sqrt{2\pi} \Theta_a \right) 
 \cos \left( \sqrt{4\pi} \Phi_s \right)$ is $2K_s + 1/(2K_a)$;
the scaling dimension of 
$\cos \left( \sqrt{2\pi} \Theta_a \right) 
 \cos \left( \sqrt{4\pi} \Phi_a \right)$ is $2K_a + 1/(2K_a)$;
Therefore, $g_1$ will grow at large distances for $K_s < 1$; 
$g_2$ will grow for $K_a < 1$; $g_3$ will grow for $K_a > 1/4$.  
The $g_5$ operator is always irrelevant, and therefore will be ignored.
However, we will keep the $g_4$ term since, though seemingly irrelevant, it 
is the most relevant operator generated by the interchain coupling which 
couples the symmetric and antisymmetric modes. We will see that the $g_4$ 
term plays a subtle role.

It is interesting to study the behavior of the $xy$ and $z$ components
of the interchain coupling individually.  First we consider the case
$J_{\perp}^{xy}=0$ and $J_{\perp}^z \neq 0$.  Therefore, only
$g_1$ and $g_2$ are nonzero.  
The $g_1$ term is relevant for $0 \leq \Delta \leq 1$.   
The $g_2$ term is relevant for $K < 1 - \frac{J_{\perp}^z}{\pi u}$ (for 
weak interchain coupling.)  Therefore, the symmetric mode is gapped in 
the entire region $0 \leq \Delta \leq 1$, while the antisymmetric
mode remains gapless in a region about $\Delta = 0$ and is gapped
outside of this region.  In the region where $\Phi_a$ is pinned,
the antisymmetric mode is in an ordered phase \cite{schulz}.

Next we consider the case $J_{\perp}^{xy} \neq 0$ and 
$J_{\perp}^z = 0$.  Therefore, only $g_3$ and $g_4$ are nonzero.  
For $0 \leq \Delta \leq 1$, the $g_3$ term is relevant while
the $g_4$ term is irrelevant.  Therefore, $g_3$ will grow while 
$g_4$ will initially decrease under the RG.
However, once $g_3$ grows to ${\cal O}$(1), we can safely say 
that $\Theta_a$ is pinned.  At this point, assuming that $g_4$ is
renormalized to $\overline{g_4}$, the term that couples the symmetric
and antisymmetric modes can be written as
\begin{eqnarray}
   & &  \overline{g_4} \int \frac{dx}{(2\pi a)^2} 
    \left\langle \cos \left( \sqrt{2\pi} \Theta_a \right) \right\rangle
    \cos \left( \sqrt{8\pi} \Phi_s \right)  \nonumber \\
   & \approx &  \overline{g_4} 
     \left\langle \cos \left( \sqrt{2\pi} \Theta_a \right) \right\rangle 
     \int \frac{dx}{ (2\pi a)^2} \cos \left( \sqrt{8\pi} \Phi_s \right) \, .
\end{eqnarray}

\end{multicols}

\begin{table}
\caption{Results for the two-leg ladder 
in the region $0 \leq \Delta \leq 1$ (i.e. $1 \geq K \geq 1/2$).}
\begin{center}
\begin{tabular}{cccc}
 & $J_{\perp}^{xy} = 0$, $J_{\perp}^z \neq 0$ \ \ \ \
 & $J_{\perp}^{xy} \neq 0$, $J_{\perp}^z = 0$ \ \ \ \
 & $J_{\perp}^z / J_{\perp}^{xy} = \Delta$  \\
\tableline \\
 $\Phi_s$ & pinned & pinned & pinned \\   \\
 $\Phi_a$, $\Theta_a$ & \ \ \ $\Phi_a$ pinned for $K_a < 1$, 
   & $\Theta_a$ pinned,  & $\Theta_a$ pinned,  \\ 
       & ordered & disordered & disordered \\ \\
       & Gapless for $K_a < 1$, &               &               \\
 Phase & Haldane for $K_a < 1$  & Haldane phase & Haldane phase \\
\tableline
\end{tabular}
\end{center}
\label{table:twoleg}
\end{table}

\begin{multicols}{2}
\noindent
The expectation value is taken with respect to 
\begin{eqnarray}
 H_a & = & \frac{u_a}{2} \int dx  \left[ \overline{K_a} \Pi_a^2  +  
   \frac{1}{\overline{K_a}} (\partial_x \Phi_a)^2 \right]  \nonumber \\
    & + & \overline{g_3} \int {dx \over (2 \pi a)^2} 
        \cos \left( \sqrt{2\pi} \Theta_a \right)  \, ,
\end{eqnarray}
where $\overline{K_a}$ and $\overline{g_3}$ are the renormalized values
of $K_a$ and $g_3$.  Then we can define a new coupling constant
\begin{equation}
 g'_4 = \overline{g_4} \left\langle \cos 
        \left( \sqrt{2\pi} \Theta_a \right) \right\rangle \, .
\end{equation}
Therefore, our effective Hamiltonian for the symmetric mode is
\begin{eqnarray}
   H_{\rm eff} & = & \frac{u_s}{2} \int dx \left[ \overline{K_s} \Pi_s^2  + 
     \frac{1}{ \overline{K_s} } (\partial_x \Phi_s)^2 \right] \nonumber \\
    & + & g'_4 \int \frac{dx}{ (2\pi a)^2} \cos \left( \sqrt{8\pi} 
      \Phi_s \right)  
\end{eqnarray}
where $\overline{K_s}$ is the renormalized value of $K_s$.
Now $\cos \left( \sqrt{8\pi} \Phi_s \right)$ is relevant for 
 $\overline{K_s} < 1$.  Therefore, the term is relevant.  
We see that $J_{\perp}^{xy}$ alone generates a gap in both  
the symmetric and antisymmetric modes.
However, since $\Theta_a$ was pinned, the antisymmetric mode
is in a disordered phase. \cite{schulz}

Finally, we consider the case $J_{\perp}^{xy} \neq 0$ and 
$J_{\perp}^z \neq 0$ with $J_{\perp}^z/J_{\perp}^{xy} = \Delta$.
Therefore, for $J_{\perp}^{xy} = 1$, we recover the results for 
the anisotropic spin-1 chain.  
The $g_1$ and $g_3$ terms are relevant in the entire range 
$0 \leq \Delta \leq 1$, and $g_3$ is always more relevant than $g_2$.  
We see that both the symmetric and antisymmetric modes will get 
gapped, with the antisymmetric mode being in a disordered phase; the 
Haldane gap is generated for {\it arbitrarily small} interchain coupling.

The results for the two-leg ladder are summarized in 
Table \ref{table:twoleg}.


\section{ The Four-Leg Ladder }

Next we study the opening of the Haldane gap in the four-leg ladder.
The Hamiltonian is 
\begin{eqnarray}
   H & = & H_1 + H_2 + H_3 + H_4 \nonumber \\
     & &  +  H_{1,2} +  H_{1,3} + H_{1,4} +  H_{2,3} + H_{2,4} + 
       H_{3,4} \,. 
\end{eqnarray}

Let us introduce ${\bf \Phi}^T = ( \Phi_1 , \Phi_2 , \Phi_3 , \Phi_4 )$ 
and ${\bf \Pi}^T = ( \Pi_1 , \Pi_2 , \Pi_3 , \Pi_4 )$.  Then in its
bosonized form our Hamiltonian is 
\breakon
\begin{eqnarray}
 H  & = & \int dx \left[ \frac{uK}{2} {\bf \Pi}^T {\bf \Pi} \, 
 +  (\partial_x {\bf \Phi})^T {\bf M} (\partial_x {\bf \Phi}) \right] 
      \nonumber \\
   & & +  g_1 \int {dx \over (2 \pi a)^2} \left[  \cos\left( \sqrt{4\pi} 
     (\Phi_2 + \Phi_1) \right)
   +  \cos\left( \sqrt{4\pi} (\Phi_3 + \Phi_1) \right)  +   
 \cos\left( \sqrt{4\pi} (\Phi_4 + \Phi_1) \right) \right.  \nonumber \\ 
  & & \phantom{+++} + \left. \cos\left( \sqrt{4\pi} (\Phi_3 + \Phi_2) \right) 
  +  \cos\left( \sqrt{4\pi} (\Phi_4 + \Phi_2) \right)  + 
 \cos\left( \sqrt{4\pi} (\Phi_4 + \Phi_3) \right) \right] \nonumber \\
 & & + \ g_2 \int {dx \over (2 \pi a)^2} \left[ \cos\left( \sqrt{4\pi} 
    (\Phi_2 - \Phi_1) \right) 
  + \cos\left( \sqrt{4\pi} (\Phi_3 - \Phi_1) \right)   +   
    \cos\left( \sqrt{4\pi} (\Phi_4 - \Phi_1) \right) \right.  \nonumber \\ 
  & & \phantom{+++} + \left. \cos\left( \sqrt{4\pi} (\Phi_3 - \Phi_2) \right)  
   + \cos\left( \sqrt{4\pi} (\Phi_4 - \Phi_2) \right)  + 
 \cos\left( \sqrt{4\pi} (\Phi_4 - \Phi_3) \right) \right] \nonumber \\
 & & + \ g_3 \int {dx \over (2 \pi a)^2} \left[  \cos\left( \sqrt{\pi} 
    (\Theta_2 - \Theta_1) \right) 
    +  \cos\left( \sqrt{\pi} (\Theta_3 - \Theta_1) \right) +   
   \cos\left( \sqrt{\pi} (\Theta_4 - \Theta_1) \right) \right.  \nonumber \\ 
 & & \phantom{+++} +  \left. 
   \cos\left( \sqrt{\pi} (\Theta_3 - \Theta_2) \right) 
    + \cos\left( \sqrt{\pi} (\Theta_4 - \Theta_2) \right) + 
 \cos\left( \sqrt{4\pi} (\Theta_4 - \Theta_3) \right)  \right] \nonumber \\ 
 & & + \ g_4 \int {dx \over (2 \pi a)^2} \left[ 
 \cos\left( \sqrt{\pi} (\Theta_2 - \Theta_1) \right)
 \cos\left( \sqrt{4\pi} (\Phi_2 + \Phi_1) \right) 
 + \cos\left( \sqrt{\pi} (\Theta_3 - \Theta_1) \right)
   \cos\left( \sqrt{4\pi} (\Phi_3 + \Phi_1) \right)  \right. \nonumber \\ 
 & & \phantom{+++} + \left.  
   \cos\left( \sqrt{\pi} (\Theta_4 - \Theta_1) \right)   
   \cos\left( \sqrt{4\pi} (\Phi_4 + \Phi_1) \right) 
 + \cos\left( \sqrt{\pi} (\Theta_3 - \Theta_2) \right)
  \cos\left( \sqrt{4\pi} (\Phi_3 + \Phi_2) \right)  \right.  \nonumber \\
 & & \phantom{+++} + \left.
    \cos\left( \sqrt{\pi} (\Theta_4 - \Theta_2) \right) 
    \cos\left( \sqrt{4\pi} (\Phi_4 + \Phi_2) \right)  
 +  \cos\left( \sqrt{\pi} (\Theta_4 - \Theta_3) \right)
    \cos\left( \sqrt{4\pi} (\Phi_4 + \Phi_3) \right) \right] \nonumber \\
 & & + \ g_5 \int {dx \over (2 \pi a)^2} \left[ 
 \cos\left( \sqrt{\pi} (\Theta_2 - \Theta_1) \right)
 \cos\left( \sqrt{4\pi} (\Phi_2 - \Phi_1) \right) 
 + \cos\left( \sqrt{\pi} (\Theta_3 - \Theta_1) \right)
   \cos\left( \sqrt{4\pi} (\Phi_3 - \Phi_1) \right) \right.  \nonumber \\
 & & \phantom{+++} + \left.   
   \cos\left( \sqrt{\pi} (\Theta_4 - \Theta_1) \right)   
   \cos\left( \sqrt{4\pi} (\Phi_4 - \Phi_1) \right) 
 + \cos\left( \sqrt{\pi} (\Theta_3 - \Theta_2) \right)
  \cos\left( \sqrt{4\pi} (\Phi_3 - \Phi_2) \right)  \right.  \nonumber \\
 & & \phantom{+++} + \left. 
    \cos\left( \sqrt{\pi} (\Theta_4 - \Theta_2) \right) 
    \cos\left( \sqrt{4\pi} (\Phi_4 - \Phi_2) \right)  
 +  \cos\left( \sqrt{\pi} (\Theta_4 - \Theta_3) \right)
    \cos\left( \sqrt{4\pi} (\Phi_4 - \Phi_3) \right) \right] \, ,
\end{eqnarray}
\breakoff

\noindent
where {\bf M} is given by 
\begin{equation}
 {\bf M} \ = \ \left( \begin{array}{cccc}
 a & b & b & b \\
 b & a & b & b \\
 b & b & a & b \\ 
 b & b & b & a \end{array} \right)
\end{equation}
with $a = u/ (2K)$ and $b = J_{\perp}^z / \pi$.  
{\bf M} is diagonalized by the orthogonal matrix {\bf S}
\begin{equation}
 {\bf S} \ = \ \left( \begin{array}{rrrr}
 \frac{1}{2} & \frac{1}{2} & \frac{1}{2} & \frac{1}{2} \\
 \frac{1}{2} & \frac{1}{2} & -\frac{1}{2} & -\frac{1}{2} \\
 \frac{1}{2} & -\frac{1}{2} & -\frac{1}{2} & \frac{1}{2} \\
 \frac{1}{2} & -\frac{1}{2} & \frac{1}{2} & -\frac{1}{2} \end{array} \right)
 \, .
\end{equation}
It should be noted that the choice of {\bf S} is not unique.

Define ${\bf \Lambda}$ by
\begin{equation}
 {\bf \Lambda} \ = \ {\bf S}^T {\bf M} {\bf S} \, .
\end{equation}
 ${\bf \Lambda}$ is given by
\begin{equation}
 {\bf \Lambda} \ = \ \left( \begin{array}{cccc}
 a+3b & 0 & 0 & 0 \\
 0 & a - b & 0 & 0 \\
 0 & 0 & a - b & 0 \\
 0 & 0 & 0 & a - b  \end{array} \right) \, .
\end{equation}
Define the fields $\Phi_s$, $\Phi_{a1}$, $\Phi_{a2}$, and $\Phi_{a3}$ by
\begin{equation}
\begin{array}{cccc} 
\left( \begin{array}{c} \Phi_1 \\ \Phi_2 \\ 
                    \Phi_3 \\ \Phi_4 \end{array} \right) $ = $
\left( \begin{array}{rrrr}
 \frac{1}{2} & \frac{1}{2} & \frac{1}{2} & \frac{1}{2} \\
 \frac{1}{2} & \frac{1}{2} & -\frac{1}{2} & -\frac{1}{2} \\
 \frac{1}{2} & -\frac{1}{2} & -\frac{1}{2} & \frac{1}{2} \\
 \frac{1}{2} & -\frac{1}{2} & \frac{1}{2} & -\frac{1}{2} 
       \end{array} \right)
\left( \begin{array}{c} \Phi_s \\ \Phi_{a1} \\
                    \Phi_{a2} \\ \Phi_{a3}  \end{array} \right) \, . 
\end{array}
\end{equation} 
In terms of the old fields, the new fields are given by
\begin{equation}
\begin{array}{lcc}
\Phi_s & = & \frac{1}{2} \left( \Phi_1 \, + \, \Phi_2 \, + \,
 \Phi_3 \, + \, \Phi_4 \right) \\
\Phi_{a1} & = & \frac{1}{2} \left( \Phi_1 \, + \, \Phi_2 \, - \,
 \Phi_3 \, - \, \Phi_4 \right) \\
\Phi_{a2} & = & \frac{1}{2} \left( \Phi_1 \, - \, \Phi_2 \, - \,
 \Phi_3 \, + \, \Phi_4 \right) \\
\Phi_{a3} & = & \  \frac{1}{2} \left( \Phi_1 \, - \, \Phi_2 \, + \,
 \Phi_3 \, - \, \Phi_4 \right) \, .
\end{array}
\end{equation}
We see that $\Phi_s$ is a symmetric mode while $\Phi_{a1}$, $\Phi_{a2}$, and
$\Phi_{a3}$ are antisymmetric modes.   

Let us define ${\bf \tilde{\Phi}}^T  = (\Phi_s, \Phi_{a1}, \Phi_{a2}, 
\Phi_{a3})$ and ${\bf \tilde{\Pi}}^T = (\Pi_s, \Pi_{a1}, \Pi_{a2}, \Pi_{a3})$.
Then ${\bf \Phi} = {\bf S} {\bf \tilde{\Phi}}$ and
  ${\bf \Pi} = {\bf S} {\bf \tilde{\Pi}}$.  In terms of the new
fields, the quadratic part of the Hamiltonian is given by
\begin{equation}
 H_0 = \int dx \left[ \frac{uK}{2} {\bf \tilde{\Pi}}^T {\bf \tilde{\Pi}} \, 
 + \, (\partial_x {\bf \tilde{\Phi}})^T 
 {\bf \Lambda} (\partial_x {\bf \tilde{\Phi}})  \right] \,.
\end{equation}
We can write this as
\begin{eqnarray}
 H_0 & = & \frac{u_s}{2}\int dx \left[ K_s \Pi_s^2 
  + \frac{1}{K_s} (\partial_x \Phi_s)^2 \right]  \nonumber \\
  & + & \sum_{i=1}^3 \frac{u_a}{2}\int dx \left[ K_a \Pi_{ai}^2 
  +  \frac{1}{K_a} (\partial_x \Phi_{ai})^2 \right] 
\end{eqnarray}
with 
\begin{eqnarray}
 u_s = u \left( 1  +  \frac{6 K J_{\perp}^z }{u \pi} \right)^{1/2} \,, &
  & \quad
 K_s = K \left( 1  + \frac{6 K J_{\perp}^z }{u \pi} \right)^{-1/2} \,,
\nonumber \\
 & & \\
 u_a = u \left( 1  - \frac{2 K J_{\perp}^z }{u \pi} \right)^{1/2} \,, &
  & \quad 
 K_a  = K \left( 1 - \frac{2 KJ _{\perp}^z }{u \pi} \right)^{-1/2}  \,.
\nonumber 
\end{eqnarray}
For $J_{\perp}^{xy}, J_{\perp}^z \ll 1$ we have
\begin{eqnarray}
 u_s \approx u \left( 1 +  \frac{3KJ_{\perp}^z }{u \pi} \right) \,, &
 \ \ \ &
 K_s \approx K \left( 1 - \frac{3KJ_{\perp}^z }{u \pi} \right) \,,
\nonumber \\
 u_a \approx u \left( 1 -  \frac{KJ_{\perp}^z }{u \pi} \right) \,,  &
 \ \ \ &
 K_a \approx K \left( 1 +  \frac{KJ_{\perp}^z }{u \pi} \right) \,.
\end{eqnarray}

In terms of these new fields, the interchain coupling is given by
\breakon
\begin{eqnarray}
 & & 2 g_1 \int {dx \over (2 \pi a)^2} \,  
 \left[ \cos\left( \sqrt{4\pi} \Phi_{a1} \right)  
 \cos\left( \sqrt{4\pi} \Phi_s \right) \right. 
    +  \cos\left( \sqrt{4\pi} \Phi_{a2} \right)
 \cos\left( \sqrt{4\pi} \Phi_s \right) 
    + \left. \cos\left( \sqrt{4\pi} \Phi_{a3} \right)  
 \cos\left( \sqrt{4\pi} \Phi_s \right) \right]  \nonumber \\ 
 & & + 2 g_2 \int {dx \over (2 \pi a)^2}  
  \left[  \cos\left( \sqrt{4\pi} \Phi_{a2} \right)
 \cos\left( \sqrt{4\pi} \Phi_{a1} \right)   \right. 
  +  \cos\left( \sqrt{4\pi} \Phi_{a3} \right) 
 \cos\left( \sqrt{4\pi} \Phi_{a1} \right) 
  +  \left. \cos\left( \sqrt{4\pi} \Phi_{a3} \right) 
 \cos\left( \sqrt{4\pi} \Phi_{a2} \right) \right] \nonumber \\ 
 & & + 2 g_3 \int {dx \over (2 \pi a)^2}  
  \left[  \cos\left( \sqrt{\pi} \Theta_{a2} \right)
 \cos\left( \sqrt{\pi} \Theta_{a1} \right) \right. 
  +  \cos\left( \sqrt{\pi} \Theta_{a3} \right)
 \cos\left( \sqrt{\pi} \Theta_{a1} \right) 
  + \left.  \cos\left( \sqrt{\pi} \Theta_{a3} \right) 
 \cos\left( \sqrt{\pi} \Theta_{a2} \right)   \right]  \nonumber \\
 & & + \ g_4 \int {dx \over (2 \pi a)^2} \left[ 
   \cos\left( \sqrt{\pi} (\Theta_{a3} + \Theta_{a2}) \right)
   \cos\left( \sqrt{4\pi} (\Phi_s + \Phi_{a1}) \right) 
 + \cos\left( \sqrt{\pi} (\Theta_{a3} - \Theta_{a2}) \right)
   \cos\left( \sqrt{4\pi} (\Phi_s - \Phi_{a1}) \right)  \right. \nonumber \\
 & & \phantom{+++} + \left.
   \cos\left( \sqrt{\pi} (\Theta_{a2} + \Theta_{a1}) \right)
   \cos\left( \sqrt{4\pi} (\Phi_s + \Phi_{a3}) \right)     
 + \cos\left( \sqrt{\pi} (\Theta_{a2} - \Theta_{a1}) \right) 
   \cos\left( \sqrt{4\pi} (\Phi_s - \Phi_{a3}) \right)  \right. \nonumber \\  
 & & \phantom{+++} + \left.
   \cos\left( \sqrt{\pi} (\Theta_{a3} + \Theta_{a1}) \right)   
   \cos\left( \sqrt{4\pi} (\Phi_s + \Phi_{a2}) \right) 
 + \cos\left( \sqrt{\pi} (\Theta_{a3} - \Theta_{a1}) \right)
   \cos\left( \sqrt{4\pi} (\Phi_s - \Phi_{a2}) \right)  \right]  \nonumber \\
 & & + \ g_5 \int {dx \over (2 \pi a)^2} \left[ 
   \cos\left( \sqrt{\pi} (\Theta_{a3} + \Theta_{a2}) \right)
   \cos\left( \sqrt{4\pi} (\Phi_{a3} + \Phi_{a2}) \right)
 + \cos\left( \sqrt{\pi} (\Theta_{a3} - \Theta_{a2}) \right)
   \cos\left( \sqrt{4\pi} (\Phi_{a3} - \Phi_{a2}) \right) \right. \nonumber \\
 & & \phantom{+++} + \left.
   \cos\left( \sqrt{\pi} (\Theta_{a2} + \Theta_{a1}) \right)
   \cos\left( \sqrt{4\pi} (\Phi_{a2} + \Phi_{a1}) \right)  
 + \cos\left( \sqrt{\pi} (\Theta_{a2} - \Theta_{a1}) \right) 
   \cos\left( \sqrt{4\pi} (\Phi_{a2} - \Phi_{a1}) \right) \right. \nonumber \\
 & & \phantom{+++} + \left.
   \cos\left( \sqrt{\pi} (\Theta_{a1} + \Theta_{a3}) \right)   
   \cos\left( \sqrt{4\pi} (\Phi_{a1} + \Phi_{a3}) \right) 
 + \cos\left( \sqrt{\pi} (\Theta_{a3} - \Theta_{a1}) \right)
  \cos\left( \sqrt{4\pi} (\Phi_{a3} - \Phi_{a1}) \right)  \right] \, .
\end{eqnarray}
\breakoff

\noindent
We see that $g_1$ and $g_4$ couples the symmetric mode to the antisymmetic
modes, while $g_2$, $g_3$, and $g_5$ only couple the antisymmetric modes 
among themselves.

Just like for the two-leg ladder, we would like to know the relevance
or irrelevance of the operators in the interchain coupling.  Therefore, 
we consider the scaling dimension of the operators in the interchain
coupling.  The $g_1$ terms have scaling dimension $K_s + K_a$;
the $g_2$ terms have scaling dimension $2K_a$;
the $g_3$ terms have scaling dimension $1/(2K_a)$;
the $g_4$ terms have scaling dimension $K_s + K_a + 1/(2K_a)$;
the $g_5$ terms have scaling dimension $2K_a + 1/(2K_a)$.
Therefore, the $g_1$ terms are relevant for $K_s + K_a < 2$;
  the $g_2$ terms are relevant for $K_a < 1$;
  the $g_3$ terms are relevant for $K_a > 1/4$.
The $g_5$ terms are always irrelevant and will be dropped.
However, we will keep the $g_4$ terms.  Though they are irrelevant,
they are the most relevant terms (arising from the $xy$ part of the
interchain coupling) which couple the symmetric and antisymmetric modes.
We will see that the $g_4$ terms play a subtle role.

Since the $g_1$, $g_2$, and $g_3$ terms are all relevant and all of the 
modes are coupled together, the physics is determined by the operators
which reach strong coupling first under the RG.  The RG equations for
the parameters are
\begin{eqnarray}
 \frac{dg_1}{dl} & = & \left[ 2 - (K_s + K_a) \right]g_1  \nonumber \\
 \frac{dg_2}{dl} & = & \left( 2 - 2K_a \right)g_2  \nonumber \\
 \frac{dg_3}{dl} & = & \left( 2 - \frac{1}{2K_a} \right)g_3  \nonumber \\
 \frac{dg_4}{dl} & = & \left[ 2 - (K_s + K_a + \frac{1}{2K_a}) \right]g_4  
   \nonumber \\
 \frac{dK_s}{dl} & = & -6g_1^2 \left( \frac{K_s}{4\pi u_s} \right)^2  
                       -3g_4^2 \left( \frac{K_s}{4\pi u_s} \right)^2
   \nonumber \\
 \frac{dK_a}{dl} & = & -2g_1^2 \left( \frac{K_a}{4\pi u_a} \right)^2
                       -4g_2^2 \left( \frac{K_a}{4\pi u_a} \right)^2
		   + g_3^2 \left( \frac{1}{4\pi u_a} \right)^2 \nonumber \\
               & & - g_4^2 \left( \frac{K_a}{4\pi u_a} \right)^2
	    + \frac{1}{2}g_4^2 \left( \frac{1}{4\pi u_a} \right)^2 \, .
\end{eqnarray}
See Appendix~B for a derivation of the RG equations.

Just like for the two-leg ladder, it is interesting to study 
the behavior of the $xy$ and $z$ components of the interchain coupling
individually.  First we consider the case $J_{\perp}^{xy} = 0$ and 
$J_{\perp}^z \neq 0$.  For this case the initial values of the 
coupling constants are $g_1(l=0) = 4 J_{\perp}^z$, 
 $g_2(l=0) = -4J_{\perp}^z$, $g_3(l=0) = 0$, and $g_4(l=0) = 0$.
The $g_1$ terms are relevant for $0 \leq \Delta \leq 1$.  The 
$g_2$ terms are relevant for $K < 1 - \frac{J_{\perp}^z}{\pi u}$
(for weak interchain coupling.)  However, since all of the modes are coupled
(and hence locked together), both the symmetric and antisymmetric 
modes are gapped in the entire region $0 \leq \Delta \leq 1$.  Notice 
that since the $\Phi_{ai}$ were pinned, the antisymmetric modes are 
in an ordered phase.

Next we consider the case $J_{\perp}^{xy} \neq 0$ and 
 $J_{\perp}^z = 0$.  For this case the initial values of the 
coupling constants are $g_1(l=0) = 0$, $g_2(l=0) = 0$,
 $g_3(l=0) = -4\pi J_{\perp}^{xy}$, $g_4(l=0) = 2\pi J_{\perp}^{xy}$,
 $K_s(l=0) = K$, $K_a(l=0) = K$, $u_s = u$, and $u_a = u$.  The $g_3$ terms 
are relevant and the $g_4$ terms are irrelevant.  Therefore, $g_3$ will
grow and $g_4$ will initially decrease under the RG.  However, once 
$g_3 = {\cal O}(1)$ we can safely say that the $\Theta_{ai}$ are pinned.
Then we can write the $g_4$ term as
\breakon
\begin{eqnarray}
 &  & \ \overline{g_4} \int {dx \over (2 \pi a)^2} \left[ 
 \left\langle \cos\left( \sqrt{\pi} \left( \Theta_{a2} 
              + \Theta_{a3} \right) \right) \right\rangle
 \cos\left( \sqrt{4\pi} (\Phi_s + \Phi_{a1}) \right) 
 + \left\langle \cos\left( \sqrt{\pi} \left(\Theta_{a1} 
              + \Theta_{a2} \right) \right) \right\rangle
   \cos\left( \sqrt{4\pi} (\Phi_s + \Phi_{a3}) \right)  \right. \nonumber \\ 
 & & \phantom{+++} + \left.  
   \left\langle \cos\left( \sqrt{\pi} \left(\Theta_{a1} 
              + \Theta_{a3} \right) \right) \right\rangle   
   \cos\left( \sqrt{4\pi} (\Phi_s + \Phi_{a2}) \right) 
 + \left\langle \cos\left( \sqrt{\pi} \left(\Theta_{a3} 
              - \Theta_{a1} \right) \right) \right\rangle
  \cos\left( \sqrt{4\pi} (\Phi_s - \Phi_{a2}) \right)  \right.  \nonumber \\
 & & \phantom{+++} + \left.
   \left\langle \cos\left( \sqrt{\pi} \left(\Theta_{a2} 
              - \Theta_{a1} \right) \right) \right\rangle 
   \cos\left( \sqrt{4\pi} (\Phi_s - \Phi_{a3}) \right)  
 + \left\langle \cos\left( \sqrt{\pi} \left(\Theta_{a2} 
              - \Theta_{a3} \right) \right) \right\rangle
   \cos\left( \sqrt{4\pi} (\Phi_s - \Phi_{a1}) \right) \right] \nonumber \\
 & \approx & \overline{g_4} 
 \left\langle \cos \left( \sqrt{\pi} (\Theta_{a1} 
     + \Theta_{a2}) \right) \right\rangle \int {dx \over (2 \pi a)^2} \left[ 
 \cos\left( \sqrt{4\pi} (\Phi_s + \Phi_{a1}) \right) +
   \cos\left( \sqrt{4\pi} (\Phi_s + \Phi_{a3}) \right)  \right. \nonumber \\ 
 & & \phantom{+++} + \left.  
   \cos\left( \sqrt{4\pi} (\Phi_s + \Phi_{a2}) \right) + 
  \cos\left( \sqrt{4\pi} (\Phi_s - \Phi_{a2}) \right)  
   \cos\left( \sqrt{4\pi} (\Phi_s - \Phi_{a3}) \right)  + 
   \cos\left( \sqrt{4\pi} (\Phi_s - \Phi_{a1}) \right) \right] \, ,
\end{eqnarray}
where $\overline{g_4}$ is the renormalized coupling.  The expectation
value is taken with respect to
\begin{eqnarray}
 H_a & = & \sum_{i=1}^3 \frac{u_a}{2}\int dx 
       \left[ \overline{K_a} \Pi_{ai}^2 
    + \frac{1}{\overline{K_a}} (\partial_x \Phi_{ai})^2 \right]  \nonumber \\
 & + & 2 \overline{g_3} \int {dx \over (2 \pi a)^2}  
  \left[  \cos\left( \sqrt{\pi} \Theta_{a2} \right)
 \cos\left( \sqrt{\pi} \Theta_{a1} \right) \right. 
  +  \cos\left( \sqrt{\pi} \Theta_{a3} \right)
 \cos\left( \sqrt{\pi} \Theta_{a1} \right) 
  + \left.  \cos\left( \sqrt{\pi} \Theta_{a3} \right) 
 \cos\left( \sqrt{\pi} \Theta_{a2} \right)   \right] \, , 
\end{eqnarray}
where $\overline{K_a}$ and $\overline{g_3}$ are the renormalized values 
of $K_a$ and $g_3$.  Notice that $H_a$ is invariant under 
$\Theta_{ai} \rightarrow - \Theta_{ai}$;
therefore, $\langle \cos \left(\sqrt{\pi} (\Theta_{ai} - 
    \Theta_{aj}) \right) \rangle = 
    \langle \cos \left(\sqrt{\pi} (\Theta_{ai} + 
    \Theta_{aj}) \right) \rangle\, .$ 
Now we can define a new effective coupling 
\begin{equation}
 g'_4 = \overline{g_4} \left\langle \cos 
 \left(\sqrt{\pi} (\Theta_{a1} + \Theta_{a2}) \right) \right\rangle
\end{equation}
and write the $g_4$ term as
\begin{equation}
 H_1  =  2 g'_4 \sum_{i=1}^3 \int {dx \over (2 \pi a)^2} 
            \cos \left(\sqrt{4\pi} 
   \Phi_s \right)  \cos \left(\sqrt{4\pi} \Phi_{ai} \right)  \, .
\end{equation}

Recall that $\left[\Phi(x), \Theta(y) \right] = {\cal O}(1)$. 
Since the $\Theta_{ai}$ are pinned, the
 $\Phi_{ai}$ will fluctuate wildly and $e^{i\beta \Phi_{ai}}$ 
will have exponentially decaying correlations.  Looking at the
interchain coupling, one might think that the $g'_4$ term 
is now irrelevant due to the presence of the 
$\cos (\sqrt{4\pi}\Phi_{ai})$ terms.  However, this is not entirely 
correct.  To see what can happen we evaluate the (normalized)
partition function perturbatively in $g'_4$.  
We write 
\begin{eqnarray}
 \frac{Z}{Z_0} \ & = & \ \frac{1}{Z_0} \int {\cal D}\Phi e^{-S_0} e^{-S_1}  
 \nonumber  \\
     & = & \ \left\langle e^{-S_1} \right\rangle_0  \nonumber \\
 & = & 1 - \langle S_1 \rangle_0 + \frac{1}{2} \langle S_1^2 \rangle_0
 + \cdots  
\end{eqnarray}
where 
\begin{eqnarray}
 S_0 & = & \frac{u_s}{2\overline{K_s}} \int d^2x 
 \left[ \frac{1}{u_s^2} (\partial_{\tau}\Phi_s)^2 
        + (\partial_x \Phi_s)^2 \right] +
 \sum_{i=1}^3 \frac{u_a\overline{K_a}}{2} \int d^2x 
 \left[ \frac{1}{u_a^2}(\partial_{\tau}\Theta_{ai})^2 
        + (\partial_x \Theta_{ai})^2 \right]
 \nonumber \\
 & + & 2 \overline{g_3} \int {d^2x \over (2 \pi a)^2}  
  \left[  \cos\left( \sqrt{\pi} \Theta_{a2} \right)
 \cos\left( \sqrt{\pi} \Theta_{a1} \right) \right. 
  +  \cos\left( \sqrt{\pi} \Theta_{a3} \right)
 \cos\left( \sqrt{\pi} \Theta_{a1} \right) 
 + \left. \cos\left( \sqrt{\pi} \Theta_{a3} \right) 
 \cos\left( \sqrt{\pi} \Theta_{a2} \right) \right]  
\end{eqnarray} 
with $\overline{K_s}$ the renormalized value of $K_s$,
$\langle \ \rangle_0$ denotes averaging with respect to $S_0$,
and
\begin{equation}
 S_1  =  2g'_4 \sum_{i=1}^3 \int {d^2x \over (2 \pi a)^2} 
         \cos \left(\sqrt{4\pi} \Phi_s \right)  
         \cos \left(\sqrt{4\pi} \Phi_{ai} \right)  \, .
\end{equation} 
The first non-vanishing correction is
\begin{eqnarray}
 \langle S_1^2 \rangle_0 = 3 (2\overline{g_4})^2 \int  
    {d^2x_1 \over (2 \pi a)^2} {d^2x_2 \over (2 \pi a)^2} & &   
    \left\langle   
       \cos \left(\sqrt{4\pi} \Phi_s(x_1) \right)
       \cos \left(\sqrt{4\pi} \Phi_s(x_2) \right) 
       \cos \left(\sqrt{4\pi} \Phi_{a1}(x_1) \right)
       \cos \left(\sqrt{4\pi} \Phi_{a1}(x_2) \right)
     \right\rangle_0 \, .
\end{eqnarray}
\breakoff
\noindent
(We have ignored correlations in fluctuations between  
$\Phi_{ai}$ and $\Phi_{aj}$ for $i \neq j$, since they are suppressed.)
Since   $e^{i\beta \Phi_{ai}}$  have exponentially decaying
correlations,  we approximate \cite{schulz}
\begin{eqnarray}
    \lefteqn{ \left\langle  
 \cos \left(\sqrt{4\pi} \Phi_{a1}(x_1) \right)
 \cos \left(\sqrt{4\pi} \Phi_{a1}(x_2) \right) 
    \right\rangle_0 } \nonumber \\
    & & \approx  {\rm D}~\delta(x_1 - x_2)  + {\rm higher \,order \,
    corrections} \, ,
\end{eqnarray}
where D is an ${\cal O}(1)$ constant.  Therefore, we have
\begin{eqnarray}
 \frac{Z}{Z_0} \ & = & \ 1 \, + \, \frac{3}{2} (2g'_4)^2~{\rm D}  
     \int {d^2x_1 \over (2 \pi a)^2} 
     \left\langle \cos^2 \left( \sqrt{4\pi} \Phi_s \right) 
     \right\rangle_0    
 + \cdots
  \nonumber \\
 & & \\
  & = &  1 \, + \, 3 (g'_4)^2~{\rm D} \int 
         {d^2x_1 \over (2 \pi a)^2} \left\langle 
        \cos \left( 2\sqrt{4\pi} \Phi_s \right) 
        \right\rangle_0 + \cdots \, . \nonumber
\end{eqnarray}
Re-exponentiating, we have
\begin{equation}
 \frac{Z}{Z_0} \ = \ \langle e^{-S_1'} \rangle_0 
\end{equation}
where
\begin{equation}
 S_1' \ = \ g \int \frac{d^2x}{(2\pi a)^2} 
          \cos (\sqrt{16\pi} \Phi_s) \, .
\end{equation}

\end{multicols}

\vspace{.2in}
\begin{table}
\caption{Results for the four-leg ladder
 in the region $0 \leq \Delta \leq 1$ (i.e. $1 \geq K \geq 1/2$).}
\begin{center}
\begin{tabular}{cccc}  
  & $J_{\perp}^{xy} = 0$, $J_{\perp}^z \neq 0$  
  & $J_{\perp}^{xy} \neq 0$, $J_{\perp}^z = 0$ 
  & $J_{\perp}^z / J_{\perp}^{xy} = \Delta$   \\  
\tableline  \\
 $\Phi_s$ & pinned for $0 \leq \Delta \leq 1$  
          & pinned for $\overline{K_s} < 1/2$ 
          & pinned for $\overline{K_s} < 1/2$ \\  \\
 $\Phi_a$, $\Theta_a$ & $\Phi_a$ pinned, & $\Theta_a$ pinned, 
  & $\Theta_a$ pinned,  \\ 
       & ordered & disordered & disordered \\  \\
       &   
       & \ \ \ Gapless  for $\overline{K_s} > 1/2$,
       & \ \ \ Gapless  for $\overline{K_s} > 1/2$,  \\
 Phase & \ \ \ Haldane  for $0 \leq \Delta \leq 1$    
       & \ \ \ Haldane  for $\overline{K_s} < 1/2$   
       & \ \ \ Haldane  for $\overline{K_s} < 1/2$ \\
\tableline
\end{tabular}
\end{center}
\label{table:fourleg}
\end{table}

\begin{multicols}{2}
\noindent
Therefore, our effective Hamiltonian for $\Phi_s$ is 
\begin{eqnarray}
 H_{\rm eff} & = & \frac{u_s}{2} \int dx \left[ \overline{K_s} \Pi_s^2 \, 
               + \, \frac{1}{ \overline{K_s} } (\partial_x \Phi_s)^2 
                \right]    \nonumber \\   
   & & + g \int {dx \over (2 \pi a)^2}  
     \cos \left( \sqrt{16\pi} \Phi_s \right) \,,
\end{eqnarray}
where we have introduced a new effective coupling constant, $g$.

Now we have an effective Hamiltonian for $\Phi_s$ which is a standard 
sine-Gordon Hamiltonian.
The scaling dimension of $\cos \left( \sqrt{16\pi} \Phi_s \right)$
is $4 \overline{K_s}$; it is relevant for $\overline{K_s} < 1/2$.
Therefore, $g$ grows strong for $\overline{K_s} < 1/2$; $\Phi_s$
gets pinned (and hence becomes massive) for $\overline{K_s} < 1/2$.
However, the $\Theta_{ai}$ are pinned for $0 \leq \Delta \leq 1$.  
Therefore, the antisymmetric modes are in a disordered phase for 
$0 \leq \Delta \leq 1$.  

Finally, we consider the case $J_{\perp}^{xy} \neq 0$ and 
$J_{\perp}^z \neq 0$ with $J_{\perp}^z/J_{\perp}^{xy} = \Delta$.
Therefore, for $J_{\perp}^{xy} = 1$, we recover the results for 
the anisotropic spin-2 chain.  For this case the initial values of
the coupling constants are $g_1(l=0) = 4 J_{\perp}^z$, 
 $g_2(l=0) = -4J_{\perp}^z$, $g_3(l=0) = -4\pi J_{\perp}^{xy}$, 
and $g_4(l=0) = 2\pi J_{\perp}^{xy}$.  Looking at the RG equations, 
we see that for $1 \geq K \geq 1/2$ (i.e. for $0 \leq \Delta \leq 1$), 
the $g_3$ term will dominate; it is the most relevant operator.  
Therefore, the $\Theta_{ai}$ fields will get pinnned.  Once the 
$\Theta_{ai}$ fields are pinned, just like the  
$J_{\perp}^{xy} \neq 0$ \& $J_{\perp}^z = 0$ case, the $g_4$ term
becomes 
\begin{equation}
 2g'_4 \sum_{i=1}^3 
   \int {d^2x \over (2 \pi a)^2} \cos \left(\sqrt{4\pi} 
   \Phi_s \right) \cos \left(\sqrt{4\pi} \Phi_{ai} \right) \, ,
\end{equation}
where $g'_4 = \overline{g_4}\left\langle 
 \cos\left(\sqrt{\pi} (\Theta_1 + \Theta_2) \right) \right\rangle$,
and $\overline{g_4}$ is the renormalized value of $g_4$.
Define $g_1' = 2 ({\overline g_1} + g'_4) $.  
(${\overline g_1}$ is the renormalized value of $g_1$.)
Then our interchain coupling is 
\begin{equation}
 H_{\rm interchain} = g_1' \sum_{i=1}^3 \int {d^2x \over (2 \pi a)^2} 
   \cos \left(\sqrt{4\pi} 
   \Phi_s \right) \cos \left(\sqrt{4\pi} \Phi_{ai} \right) \, .
\end{equation}

From this point, the analysis proceeds identically to the case
 $J_{\perp}^{xy} \neq 0$ \& $J_{\perp}^z = 0$ --- derive an effective 
sine-Gordon Hamiltonian for the symmetric mode and determine the 
phase boundary from this effective Hamiltonian.  Similar to the 
case  $J_{\perp}^{xy} \neq 0$ \& $J_{\perp}^z = 0$, the effective
sine-Gordon Hamiltonian is
\begin{eqnarray}
 H_{\rm eff} & = & \frac{u_s}{2} \int dx \left[ \overline{K_s} \Pi_s^2 \, 
               + \, \frac{1}{ \overline{K_s} } (\partial_x \Phi_s)^2 
                \right]    \nonumber \\   
   & & + g \int {dx \over (2 \pi a)^2}  
     \cos \left( \sqrt{16\pi} \Phi_s \right) \,,
\end{eqnarray}
where $\overline{K_s}$ is the renormalized value of $K_s$, and $g$ is
an effective coupling constant.  Again,
$\Phi_s$ will become massive for $\overline{K_s} < 1/2$,  
and the $\Theta_{ai}$ are pinned for $0 \leq \Delta \leq 1$.
Therefore, a Haldane gap appears for $ \overline{K_s} < 1/2$ and the 
antisymmetric modes are in a disordered phase for $0 \leq \Delta \leq 1$. 

The above results for the four-leg ladder are summarized in 
Table \ref{table:fourleg}.

To determine the phase boundary between the gapless and Haldane phases 
for the case $J_{\perp}^{xy} \neq 0$, $J_{\perp}^z \neq 0$, with
$J_{\perp}^z / J_{\perp}^{xy} = \Delta$, we numerically 
integrated up the RG equations.
It turned out that $g_3$ grew so much faster than $g_1$ or $g_2$.
Our numerical procedure was the following.
We integrated up the RG equations to a scale, $\xi$.  $\xi$ was 
defined as the scale where $g_3 = {\cal O}(1)$ ($g_3 \gg g_1,g_2,g_4$.)  
At that point, we could safely say that the $\Theta_{ai}$ fields 
are pinned (up to gapped fluctuations) \cite{hsiuhau}.
Then we defined 
\begin{equation}
  \overline{K_s} \equiv  K_s(\xi).
\end{equation}
From the above arguments, $\overline{K_s} = 1/2$ defines the 
phase boundary.

\begin{figure}
\epsfxsize=3.15in
\centerline{\epsfbox{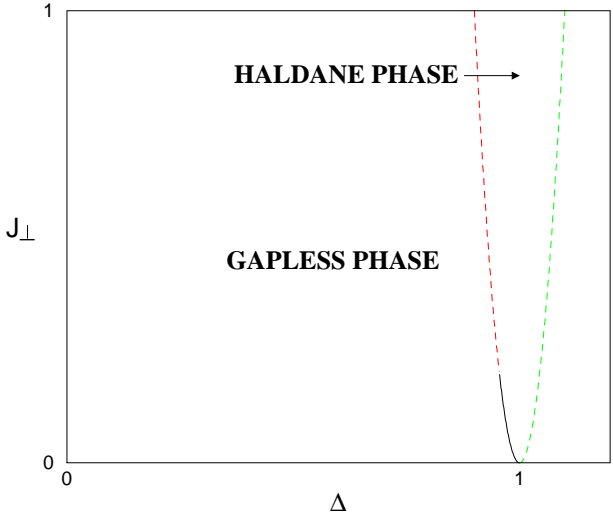}}
\caption{Phase diagram of the four-leg ladder. The solid line shows
the phase boundary between the gapless and Haldane phases calculated from
the RG equations. The dashed line indicates a speculated
phase boundary in the region where the perturbative treatment is not valid.}
\label{fig:phase-diagr}
\end{figure}

Unfortunately, we were only able to determine the phase boundary 
qualitatively.  This is because $\overline{K_s}$ is defined 
when $g_3 = {\cal O}(1)$.  (We need $g_3 = {\cal O}(1)$ so that we can 
safely say that the $\Theta_{ai}$ fields are pinned.)
However, our RG equations are no longer valid when any of the 
couplings are of ${\cal O}(1)$, since they were calculated to lowest
order for weak coupling.  Therefore, our procedure becomes
uncontrolled.  However, as long as there is no intermediate
fixed point, our results will be qualitatively correct.  Although
we cannot think of what such a fixed point would physically correspond
to, it cannot be ruled out.  The phase diagram is shown in 
Fig.~\ref{fig:phase-diagr}.

\vspace{.2in}


\section{Concluding Remarks}

In this paper we studied the generation of the Haldane gap in two-leg 
and four-leg anisotropic spin ladders using a particular form for the 
interchain coupling, namely the case where spins situated diagonally 
interact.  An interchain coupling of this form was chosen so
that when $J_{\perp} = 1$, the two-leg (four-leg) ladder was equivalent
to a spin-1 (spin-2) chain (as far as their low lying excitations are
concerned.)  

After bosonizing our model, we made a (linear) transformation to a new 
set of fields. One field contained a symmetric combination of the operators 
on the legs; the others were, in a general sense, antisymmetric combinations.
For the antisymmetric modes,
whenever $\Theta_{a}$ or $\Theta_{ai}$ were present, the operators 
containing $\Theta_a$ or $\Theta_{ai}$ were the most relevant operators;
the $\Theta_a$ or $\Theta_{ai}$ were
pinned in the entire range $0 \leq \Delta \leq 1$.  However, when only
$\Phi_a$ or $\Phi_{ai}$ were present, the behavior depended on the number 
of legs of the ladder.  For the two-leg ladder there
was a region about $\Delta=0$ where $\Phi_a$ remained gapless; for the 
four-leg ladder, the $\Phi_{ai}$ were pinned in the entire range 
$0 \leq \Delta \leq 1$.

For the symmetric mode, it was always $\Phi_s$ (rather than $\Theta_s$)
which came into play.  However, its behavior was strongly dependent on the 
number of legs of the ladder.
For the two-leg ladder the symmetric field, $\Phi_s$, was pinned in the
entire range $0 \leq \Delta \leq 1$, and the Haldane gap was generated 
for arbitrarily small interchain coupling.  
However, for the four-leg ladder, 
the Haldane gap was strongly dependent on both the interchain coupling and 
$\Delta$. In order to determine the phase diagram, the RG equations for 
the parameters were derived and integrated. It was found that the Haldane 
phase appeared in a narrow range around the isotropic point, as shown in 
Fig.~\ref{fig:phase-diagr}.  Therefore, for the two-leg ladder it is
scenario (a) of Fig.~\ref{fig:expect} which occurs, while for the four-leg
ladder it is scenario (d) of Fig.~\ref{fig:expect} which occurs; the
opening of the Haldane gap is qualitatively different for the two-leg
and four-leg ladders.

How do our results apply to spin ladders where the coupling is 
antiferromagnetic and between nearest neighbor spins on the same rung
(as in actual ladder materials)?  In Sec.~II. we argued that the 
isotropic composite spin model has the same phase diagram as the 
isotropic antiferromagnetic ladder. Now what if we include anisotropy 
and allow for the interchain coupling to be different from the 
coupling along the chains?  It is likely that the phase boundaries 
(between the massless and Haldane phases) in the $\Delta - J_{\perp}$ 
plane, which we found using the composite spin model, will shift. 
For the two-leg ladder it appears that the Haldane phase still 
opens from  $\Delta = 0, J_{\perp} = 0$ \cite{orignac}.  
However, for the four-leg ladder, it is possible that the Haldane 
phase only exists at the isotropic point or in an extremely 
narrow sliver about the isotropic point.  
We leave this (and other possibilities) for future work.


\section*{ Acknowledgements }

We would like to thank T. Giamarchi and R. Konik for clarifying discussions.
EHK gratefully acknowledges the warm hospitality of Argonne National 
Laboratory and the University of Chicago, where parts of this 
manuscript were written. EHK is also grateful to the Research
Institute for Solid State Physics (Budapest), JS to the University of
California at Santa Barbara for their warm hospitality during mutual visits.  
This work was supported by the Joint US-Hungarian Grant No. 555.
EHK also acknowledges partial support by NSF grant No. DMR-9527304.

\appendix 

\section{Bosonization Dictionary}

The bosonization procedure is well documented \cite{tsvelik,affleck}.
However, since various conventions exist in the literature, we present
our conventions here for clarity and completeness.  

To study the low energy properties of our fermion model, we linerarize
the dispersion about the Fermi points and write 
\begin{equation}
 \psi(x) = c_j / \sqrt{a} \ = \  e^{- ik_F x} \psi_R(x) 
     \ + \  e^{ik_F x} \psi_L(x) \,,
\end{equation}
where $x = ja$. 

Then we have the following bosonization rules \cite{tsvelik}
\begin{eqnarray}
 \psi_R(x) \ & = & \ \frac{1}{\sqrt{2\pi a} }
        \exp( i \sqrt{4 \pi} \phi_R )  \,,
 \nonumber \\ 
 \psi_L(x) \ & = & \ \frac{1}{\sqrt{2\pi a} }
        \exp( -i \sqrt{4 \pi} \phi_L )  \, ,
 \nonumber \\
 \psi_R^{\dagger}(x) \psi_R(x) & + & \psi_L^{\dagger}(x) \psi_L(x) 
  =  \rho_0 + \frac{1}{\sqrt{\pi} } \partial_x \Phi(x) \,.
\end{eqnarray}
where the chiral boson fields are related to the standard Bose-field, 
$\Phi$, and its dual field, $\Theta$, via
\begin{equation}
     \Phi = \phi_R + \phi_L \,, \qquad \Theta = \phi_R - \phi_L \,.
\end{equation}
The fields $\Phi$ and $\Theta$ satisfy the commutation relations
\begin{equation}
 \left[ \Phi(x), \Theta(y) \right] = i \theta(y-x) \, ,
\end{equation}
where $\theta$ is the step-function.
$\Phi$ and $\Theta$ can be thought of as order and disorder fields,
respectively, in the usual statistical mechanical sense. 
\cite{dennijs,schulz} 

Our fields are normalized such that the free boson action is given by
\begin{equation}
 S_0  =  \frac{1}{2} \int d^2x 
        \left[  \left( \partial_{\tau} \Phi \right)^2 \, + \,
         \left( \partial_x \Phi \right)^2 \right]  \,.
\label{action}
\end{equation}
The corresponding Green's function is 
\begin{equation}
 G(x,\tau) \ = \ 
  \frac{1}{4\pi} \ln \left( \frac{R^2}{\mid z \mid^2} \right) \, ,
\end{equation}
where $R$ is an infrared cutoff ($R \rightarrow \infty$).
It is the solution to
\begin{equation}
 -\left( \partial_{\tau}^2 + \partial_x^2 \right) G(x,\tau)
 \ = \ \delta(x-x_0) \delta(\tau - \tau_0)  \, .
\end{equation} 
Note that in our calculations we use the lattice spacing, $a$, 
as an ultraviolet regulator.

For a Hamiltonian of the form
\begin{equation}
  H = \frac{u}{2} \int dx 
  \left[ K \Pi^2 + \frac{1}{K} (\partial_x \Phi)^2 \right] \, ,
\end{equation}
we make a canonical transformation
\begin{equation}
  \Pi = \frac{1}{\sqrt{K}} \tilde{\Pi}, \ \ \ 
  \Phi = \sqrt{K} \tilde{\Phi}.
\label{canonical}
\end{equation}
In terms of these new variables, our Hamiltonian has the form
\begin{equation}
  H = \frac{u}{2} \int dx 
  \left[ \tilde{\Pi}^2 + (\partial_x \tilde{\Phi})^2 \right] \, .
\end{equation}
Using this form, we can pass to a Lagrangian description and obtain 
an action of the form in Eq.~(\ref{action}).

In the critical regime of the spin-1/2 Heisenberg chain, the spin-spin
correlations functions have power law behavior which can be 
calculated in the boson representation using the relations \cite{tsvelik}
\begin{eqnarray}
 \left\langle e^{i \alpha \Phi(x)} e^{-i \alpha \Phi(y)} \right\rangle 
 & = & \left( \frac{a^2}{|x-y|^2} \right)^{\frac{\alpha^2 K}{4\pi} } 
 \nonumber \\
 \left\langle e^{i \beta \Theta(x)} e^{-i \beta \Theta(y)} \right\rangle 
 & = & \left( \frac{a^2}{|x-y|^2} \right)^{\frac{\beta^2}{4\pi K}  }
 \, ,
\label{correlators}
\end{eqnarray} 
where we have first used the transformation in Eq.~(\ref{canonical}). 
From Eq.~(\ref{correlators}) we can obtain the scaling dimensions 
of the operators $e^{i \alpha \Phi(x)}$ and $e^{i \beta \Theta(x)}$;
it follows that $e^{i \alpha \Phi(x)}$ has scaling dimension 
$\frac{\alpha^2 K}{4\pi}$ and $e^{i \beta \Theta(x)}$ has scaling 
dimension $\frac{\beta^2}{4\pi K} $.  
An operator is relevant if its scaling dimension is less than 2; it is 
irrelevant if its scaling dimension is greater than 2.  If an operator 
is relevant, its coefficient grows strong at large distances; if it is 
irrelevant, its coefficient becomes smaller at large distances.  
Therefore,
 $e^{i \alpha \Phi(x)}$ is relevant for  $\frac{\alpha^2 K}{4\pi} < 2$;
 $e^{i \beta \Theta(x)}$ is relevant for $\frac{\beta^2}{4\pi K} < 2 $.


\section{Renormalization Group Analysis}

For weak interchain coupling we can analyze the physics of
our model with a perturbative RG analysis.  In an RG treatment we 
coarse grain our system, integrating out short wavelength (high energy) 
degrees of freedom, and obtain an effective theory for the long 
wavelength (low energy) degrees of freedom. 

We will derive the RG equations for $g_1$, $g_4$ and $K_s$ in 
detail to illustrate the method \cite{jose,giaschulz} .  
The RG equations for the other parameters can be derived in a similar way.

We consider the correlator  $\langle e^{i \sqrt{4\pi} \Phi_s(x_1)} 
 e^{-i \sqrt{4\pi} \Phi_s(x_2)} \rangle$ and evaluate it perturbatively 
in the interaction.  Therefore, we write
\begin{eqnarray}
  & & \left\langle e^{i \sqrt{4\pi} \Phi_s(x_1)} 
 e^{-i \sqrt{4\pi} \Phi_s(x_2)} \right\rangle  \nonumber \\
  & & =  \left\langle e^{i \sqrt{4\pi} \Phi_s(x_1)} 
 e^{-i \sqrt{4\pi} \Phi_s(x_1)} \, e^{-S_1} \right\rangle_0  \nonumber \\
 & &  =   \left\langle e^{i \sqrt{4\pi} \Phi_s(x_1)} 
 e^{-i \sqrt{4\pi} \Phi_s(x_2)} \right\rangle_0   \nonumber \\
   & & -  \left\langle e^{i \sqrt{4\pi} \Phi_s(x_1)} 
 e^{-i \sqrt{4\pi} \Phi_s(x_2)} \, S_1 \right\rangle_0 \nonumber \\
   & & + \frac{1}{2} \left\langle e^{i \sqrt{4\pi} \Phi_s(x_1)} 
 e^{-i \sqrt{4\pi} \Phi_s(x_2)} \, S_1^2 \right\rangle_0 \, + \, \cdots
\end{eqnarray}
where $\langle \ \rangle_0$ denotes averaging with respect to the
free boson action.
The first nonvanishing correction comes from the $S_1^2$ term.
\[
 \left\langle e^{i \sqrt{4\pi} \Phi_s(x_1)} 
 e^{-i \sqrt{4\pi} \Phi_s(x_2)} \, S_1^2 \right\rangle_0 \, = \,
\]
\breakon
\begin{eqnarray}
 & = & \ \frac{3}{4}  (2 g_1)^2 \int {d^2x_3 \over (2 \pi a)^2} 
     {d^2x_4 \over (2 \pi a)^2}
  \left\langle e^{i \sqrt{4\pi} \Phi_s(x_1)} 
 e^{-i \sqrt{4\pi} \Phi_s(x_2)} e^{i \sqrt{4\pi} \Phi_s(x_3)} 
 e^{-i \sqrt{4\pi} \Phi_s(x_4)} \right\rangle_0
  \left\langle e^{i \sqrt{4\pi} \Phi_{a1}(x_3)} 
 e^{-i \sqrt{4\pi} \Phi_{a1}(x_4)} \right\rangle_0  \nonumber \\
 & + & \ \frac{3}{2} g_4^2  \int {d^2x_3 \over (2 \pi a)^2} 
     {d^2x_4 \over (2 \pi a)^2}
  \left\langle e^{i \sqrt{4\pi} \Phi_s(x_1)} 
 e^{-i \sqrt{4\pi} \Phi_s(x_2)} e^{i \sqrt{4\pi} \Phi_s(x_3)} 
 e^{-i \sqrt{4\pi} \Phi_s(x_4)} \right\rangle_0
  \left\langle e^{i \sqrt{4\pi} \Phi_{a1}(x_3)} 
 e^{-i \sqrt{4\pi} \Phi_{a1}(x_4)} \right\rangle_0  \nonumber \\
 & & \hspace{2in}  \times \left\langle e^{i \sqrt{\pi} \Theta_{a2}(x_3)} 
 e^{-i \sqrt{\pi} \Theta_{a2}(x_4)} \right\rangle_0
  \left\langle e^{-i \sqrt{\pi} \Theta_{a3}(x_3)} 
 e^{i \sqrt{\pi} \Theta_{a3}(x_4)} \right\rangle_0 \, . 
\end{eqnarray}
Evaluate the correlators to get 
\begin{eqnarray}
 & = & \ \frac{3}{4} (2 g_1)^2 \int {d^2x_3 \over (2 \pi a)^2} 
    {d^2x_4 \over (2 \pi a)^2}
 \left( \frac{a^2}{\mid z_{21} \mid^2} \right)^{K_s}
 \left( \frac{\mid z_{31} \mid^2 \mid z_{42} \mid^2}
             {\mid z_{32} \mid^2 \mid z_{41} \mid^2} \right)^{K_s}
 \left( \frac{a^2}{\mid z_{43} \mid^2} \right)^{K_s}
 \left( \frac{a^2}{\mid z_{43}^a \mid^2} \right)^{K_a}  \nonumber \\
 & + & \ \frac{3}{2} g_4^2 \int {d^2x_3 \over (2 \pi a)^2} 
    {d^2x_4 \over (2 \pi a)^2}
 \left( \frac{a^2}{\mid z_{21} \mid^2} \right)^{K_s}
 \left( \frac{\mid z_{31} \mid^2 \mid z_{42} \mid^2}
             {\mid z_{32} \mid^2 \mid z_{41} \mid^2} \right)^{K_s}
 \left( \frac{a^2}{\mid z_{43} \mid^2} \right)^{K_s}
 \left( \frac{a^2}{\mid z_{43}^a \mid^2} \right)^{K_a + \frac{1}{2K_a} }
\end{eqnarray}
where $z_{ij} = u_s(\tau_i - \tau_j) + i(x_i - x_j)$ and
 $z_{ij}^a = u_a(\tau_i - \tau_j) + i(x_i - x_j)$.
Define
\begin{equation}
 {\bf r}_i = u_s \tau_i~\hat{x} +  x_i~\hat{y}
\end{equation}
and make the change of variables
\begin{equation}
  {\bf r} ={\bf r}_4 - {\bf r}_3
\ \ \ ; \ \ \
{\bf R} = \frac{1}{2} \left( {\bf r}_4 + {\bf r}_3 \right) \, .
\end{equation}
Then we have
\begin{eqnarray}
 & = & \  \frac{3}{4} (2 g_1)^2 \frac{1}{u_s^2} 
 \left( \frac{a^2}{\mid z_{21} \mid^2} \right)^{K_s} 
 \int {d^2R \over (2 \pi a)^2} {d^2r \over (2 \pi a)^2} \left[
 \frac{ \mid {\bf R} - \frac{1}{2}{\bf r} - {\bf r_1} \mid^2  
        \mid {\bf R} + \frac{1}{2}{\bf r} - {\bf r_2} \mid^2 }
      { \mid {\bf R} - \frac{1}{2}{\bf r} - {\bf r_2} \mid^2  
        \mid {\bf R} + \frac{1}{2}{\bf r} - {\bf r_1} \mid^2 } \right]^{K_s}
  \nonumber \\  & & \hspace{1.5in} \times 
  \left( \frac{a^2}{r^2} \right)^{K_s + K_a}
 \left( \frac{1}{\sin^2 \theta + 
        \left(\frac{u_a}{u_s}\right)^2 \cos^2 \theta } \right)^{K_a}
  \nonumber \\ 
 & + & \ \frac{3}{2} g_4^2   \frac{1}{u_s^2} 
 \left( \frac{a^2}{\mid z_{21} \mid^2} \right)^{K_s} 
 \int {d^2R \over (2 \pi a)^2} {d^2r \over (2 \pi a)^2} \left[
 \frac{ \mid {\bf R} - \frac{1}{2}{\bf r} - {\bf r_1} \mid^2  
        \mid {\bf R} + \frac{1}{2}{\bf r} - {\bf r_2} \mid^2 }
      { \mid {\bf R} - \frac{1}{2}{\bf r} - {\bf r_2} \mid^2  
        \mid {\bf R} + \frac{1}{2}{\bf r} - {\bf r_1} \mid^2 } \right]^{K_s}
  \nonumber \\  & & \hspace{1.5in}  \times
  \left( \frac{a^2}{r^2} \right)^{K_s+K_a+\frac{1}{2K_a}}
 \left( \frac{1}{\sin^2 \theta + 
 \left(\frac{u_a}{u_s}\right)^2 \cos^2 \theta } \right)^{K_a+\frac{1}{2K_a}} 
\end{eqnarray}
where $u_s(\tau_4 - \tau_3) = r~\cos \theta$ and $x_4 - x_3 = r~\sin \theta$.
Write 
\begin{equation}
  \left( \frac{1}{\sin^2 \theta + 
        \left(\frac{u_a}{u_s}\right)^2 \cos^2 \theta } \right)^{\beta} 
  =  1 + \beta \left[1 - \left(\frac{u_a}{u_s}\right)^2 \right] 
     \cos^2 \theta  + \cdots \,.
\end{equation}
Since $1 - \left(\frac{u_a}{u_s}\right)^2 = {\cal O}(J_{\perp}^z)$, 
the anisotropy between $u_a$ and $u_s$ gives us corrections which 
are of higher order.  Therefore, to this order we ignore the anisotropy.

Using that the most singular part of the integrand is for $r$ near $0$,
we expand the integrand about $r=0$.
We get
\begin{equation}
 \left[ \frac{ \mid {\bf R} - \frac{1}{2}{\bf r} - {\bf r_1} \mid^2  
        \mid {\bf R} + \frac{1}{2}{\bf r} - {\bf r_2} \mid^2 }
      { \mid {\bf R} - \frac{1}{2}{\bf r} - {\bf r_2} \mid^2  
        \mid {\bf R} + \frac{1}{2}{\bf r} - {\bf r_1} \mid^2 } 
 \right]^{K_s} 
 = - 4 K_s^2 r^2 \cos^2\theta  
     \frac{({\bf R} - {\bf r_1}) \cdot ({\bf R} - {\bf r_2})}
            {\mid {\bf R} - {\bf r_1} \mid^2 
             \mid {\bf R} - {\bf r_2} \mid^2} 
   + {\rm disconnected \ pieces} \, ,
\end{equation}
where we have kept only connected pieces \cite{konik,boyanovsky}.
 
Therefore we have
\[
 \left\langle e^{i \sqrt{4\pi} \Phi_s(x)} 
 e^{-i \sqrt{4\pi} \Phi_s(y)} \, S_1^2 \right\rangle_0
\]
\begin{eqnarray}
 & = &\  \left( \frac{a^2}{\mid z_{21} \mid^2} \right)^{K_s}
  \int {d^2R \over (2 \pi a)^2} 
  \nabla_{ {\bf R} } \ln \left(\frac{\mid {\bf R} - {\bf r_1} \mid}
                                     {a}\right) \cdot
  \nabla_{ {\bf R} } \ln \left(\frac{\mid {\bf R} - {\bf r_2} \mid}
                                     {a}\right) 
 \left[- 3 g_1^2 \left(\frac{2K_s^2}{u_s^2}\right)^2 
 \int {d^2r \over (2 \pi a)^2} r^2 \cos^2\theta
 \left( \frac{a^2}{r^2} \right)^{K_s + K_a} \right. \nonumber \\
  & & \left. \hspace{1.5in} - \, 
 \frac{3}{2} g_4^2  \left(\frac{2K_s^2}{u_s^2}\right)
 \int {d^2r \over (2 \pi a)^2} r^2 \cos^2\theta
 \left( \frac{a^2}{r^2} \right)^{K_s + K_a + \frac{1}{2K_a}} \right] \, ,
\end{eqnarray}
where we have used that 
\[
   \frac{({\bf R} - {\bf r})}{\mid {\bf R} - {\bf r} \mid^2} 
  \ = \ \nabla_{ {\bf R} } \ln \left(\frac{\mid {\bf R} - {\bf r} \mid}
                                     {a}\right) \, .
\]
Integrate over $\theta$, integrate by parts over ${\bf R}$ and
ignore the surface term, and use that
\[
 \nabla_{{\bf R}}^2 \ln \left( \frac{\mid {\bf R} - {\bf r} \mid}{a} \right) 
    \ = \ 2\pi \delta^2( {\bf R} - {\bf r}) 
\]
 to get
\begin{eqnarray} 
 \left\langle e^{i \sqrt{4\pi} \Phi_s(x)} 
 e^{-i \sqrt{4\pi} \Phi_s(y)} \, S_1^2 \right\rangle_0
 & = & \   \left( \frac{a^2}{\mid z_{21} \mid^2} \right)^{K_s} 
 \ln \left(\frac{\mid {\bf r_2} - {\bf r_1} \mid}{a}\right) 
 \left[ 6 g_1^2   \left(\frac{K_s}{2\pi u_s} \right)^2 
 \int \frac{dr}{a} \left( \frac{r}{a} \right)^{3 - 2(K_s + K_a)}
 \right. \nonumber \\
 & & \hspace{1.35in} \left. + \, 3 g_4^2      
 \left(\frac{K_s}{2\pi u_s} \right)^2 
 \int \frac{dr}{a} \left( \frac{r}{a} \right)^{3 - 2(K_s+K_a+\frac{1}{2K_a})}
 \right] \, .
\end{eqnarray}

Finally, we have 
\[
 \left\langle e^{i \sqrt{4\pi} \Phi_s(x)} 
 e^{-i \sqrt{4\pi} \Phi_s(y)} \, e^{-S_1} \right\rangle \ = 
\]
\begin{eqnarray}
 &  & \left( \frac{a^2}{\mid z_{21} \mid^2} \right)^{K_s} 
 \left\{ 1 \, + \,  
 \ln \left(\frac{a^2}{\mid {\bf r_2} - {\bf r_1} \mid^2}\right) 
 \left[\, - \,6 g_1^2 \left(\frac{K_s}{4\pi u_s} \right)^2 
  \int \frac{dr}{a} \left( \frac{r}{a} \right)^{3 - 2(K_s + K_a)}
 \right.\right.\nonumber \\ & & \hspace{1.5in} 
 \left.\left. \, - \, 3 g_4^2 \left( \frac{K_s}{4\pi u_s} \right)^2 
 \int \frac{dr}{a} \left( \frac{r}{a} \right)^{3 - 2(K_s+K_a+\frac{1}{2K_a})}
 \right] \right\} \, .
\end{eqnarray}
\breakoff
Treat the term in brackets as the first term in a cumulant expansion.
Re-exponentiate to get 
\begin{equation}
 \left\langle e^{i \sqrt{4\pi} \Phi_s(x)} 
 e^{-i \sqrt{4\pi} \Phi_s(y)} \, e^{-S_1} \right\rangle \ = \ 
 \left( \frac{a^2}{\mid z_{21} \mid^2} \right)^{K_s^{\rm eff}} 
\end{equation}           
where
\begin{eqnarray}
 K_s^{{\rm eff}} & = & K_s - 
  6 g_1^2 \left(\frac{K_s}{4\pi u_s}\right)^2 
   \int \frac{dr}{a} \left( \frac{r}{a} \right)^{3 - 2(K_s + K_a)} \nonumber \\
    & & - 3 g_4^2 \left(\frac{K_s}{4\pi u_s}\right)^2 
 \int \frac{dr}{a} \left( \frac{r}{a} \right)^{3 - 2(K_s+K_a+\frac{1}{2K_a})} .
\end{eqnarray}
To obtain the RG equations we split the integrals into two parts
\begin{equation}
\int_a^{\infty} \frac{dr}{a}~\left( \frac{r}{a} \right)^{\beta} 
\ = \ \int_a^{ba} \frac{dr}{a}~\left( \frac{r}{a} \right)^{\beta} 
\, + \, \int_{ba}^{\infty} \frac{dr}{a}~\left( 
          \frac{r}{a} \right)^{\beta} \, ,
\end{equation}
where $\ln~b \ll 1$. 
Perform the integral in the 1st term on the right hand side;  
rescale variables in the 2nd term so that the range of integration
runs from $a$ to $\infty$.  This gives
\begin{eqnarray}
 K_s^{\rm eff} & = & \tilde{K}_s - 
  6\tilde{g}_1^2 \left(\frac{K_s}{4\pi u_s}\right)^2 
   \int \frac{dr}{a} \left( \frac{r}{a} \right)^{3 - 2(K_s + K_a)} \nonumber \\
   & & -  3\tilde{g}_4^2 \left(\frac{K_s}{4\pi u_s}\right)^2 
   \int \frac{dr}{a} \left(\frac{r}{a}\right)^{3 - 2(K_s+K_a+\frac{1}{2K_a})} ,
\end{eqnarray}
where we have introduced
\begin{eqnarray}
 \tilde{K}_s \, & = & \, K_s \, - \, 
           6g_1^2 \left(\frac{K_s}{4\pi u_s}\right)^2 (b-1)
  \, - \,  3g_1^2 \left(\frac{K_s}{4\pi u_s}\right)^2 (b-1)
\nonumber \\
 \tilde{g}_1 \, & = & \, g_1~b^{2 - (K_s + K_a)}
 \nonumber \\
 \tilde{g}_4 \, & = & \, g_4~b^{2 - (K_s + K_a + \frac{1}{2K_a})} \, .
\end{eqnarray}
Since $\ln~b \ll 1$, we can write
 $1-b \, = \, \ln b$ and $b^x \, = \, 1 + x~\ln b$.  
Then write $\tilde{K}_s = K_s \, + \, d~K_s$, 
 $\tilde{g}_1 = g_1 \, + \, d~g_1$, $\tilde{g}_4 = g_4 \, + \, d~g_4$,
and define $dl=\ln b$.  This gives us the RG equations
\begin{eqnarray}
 \frac{d~K_s}{dl} \, & = & \, 
 -6g_1^2 \left( \frac{K_s}{4\pi u_s} \right)^2
 -3g_4^2 \left( \frac{K_s}{4\pi u_s} \right)^2 \, ,
 \nonumber \\  
 \frac{d~g_1}{dl} \, & = & \, 
                     \left[ 2 - (K_s + K_a) \right]g_1 \, ,  
 \nonumber \\
 \frac{d~g_4}{dl} \, & = & \, 
                     \left[ 2 - (K_s + K_a + \frac{1}{2K_a}) \right]g_4
 \, .
\end{eqnarray}



\end{multicols}

\begin{references}

\bibitem{dagotto}{For a review see E.\ Dagotto and T.\ M.\ Rice, Science 
    {\bf 271}, 618 (1996).}

\bibitem{haldane}{F.\ D.\ M.\ Haldane,
     Phys.\ Rev.\ Lett. {\bf 50}, 1153 (1983);
     Phys.\ Lett. {\bf 93A}, 464 (1983).}

\bibitem{schulz}{H.\ J.\ Schulz, 
         Phys.\ Rev.\ B.\ {\bf 34}, 6372, (1986).}

\bibitem{legeza}{\"O.\ Legeza, G.\ F\'ath, and J.\ S\'olyom,
      Phys.\ Rev.\ B.\ {\bf 55} 291 (1997).}

\bibitem{botet}{R.\ Botet and R.\ Jullien, 
        Phys.\ Rev.\ B {\bf 27}, 613 (1983);
     R.\ Botet, R.\ Jullien, and M.\ Kolb, 
         {\sl ibid.} {\bf 28}, 3914 (1983).}

\bibitem{timonen}{J.\ S\'olyom and J.\ Timonen,
     Phys.\ Rev.\ B {\bf 34}, 487 (1986).}

\bibitem{gomez}{G.\ G\'omez-Santos, 
        Phys.\ Rev.\ Lett.\ {\bf 63}, 790 (1989).}

\bibitem{nomura}{K.\ Nomura, 
        Phys.\ Rev.\ B {\bf 40}, 9142 (1989).}

\bibitem{sakai}{T.\ Sakai and M. Takahashi, 
        Phys.\ Rev.\ B {\bf 43}, 13 383 (1992).}

\bibitem{hida}{K.\ Hida, 
        J.\ Phys.\ Soc.\ Jpn.\ {\bf 60}, 1347 (1991).}

\bibitem{scalap}{E.\ Dagotto, J.\ Riera, and D.\ J.\ Scalapino, 
        Phys.\ Rev.\ B {\bf 45}, 5744 (1992);
     T.\ Barnes and J.\ Riera,  
        {\sl ibid.} {\bf 50}, 6817 (1994).}

\bibitem{barnes}{T.\ Barnes, E.\ Dagotto, J.\ Riera, and E.\ S.\ Swanson,
     Phys.\ Rev.\ B {\bf 47}, 3196 (1993).}

\bibitem{noack}{R.\ M.\ Noack, S.\ R.\ White, and D.\ J.\ Scalapino, 
        Phys.\ Rev.\ Lett. {\bf 73}, 882 (1994).}

\bibitem{gopalan}{S.\ Gopalan, T.\ M.\ Rice, and M.\ Sigrist, 
        Phys.\ Rev.\ B {\bf 49}, 8901 (1994);
     M.\ Sigrist, T.\ M.\ Rice, and F.\ C.\ Zhang,  
        {\sl ibid.} {\bf 49}, 12 058 (1994).}

\bibitem{watanabe}{H.\ Watanabe, K.\ Nomura, and S.\ Takada,
     J.\ Phys.\ Soc.\ Jpn.,\ {\bf 62}, 2845  (1993).}

\bibitem{legeza97}{\"O.\ Legeza and J.\ S\'olyom,
     Phys.\ Rev.\ B {\bf 56}, 14449 (1997).}

\bibitem{white}{S.\ R.\ White,
     Phys.\ Rev.\ Lett.\ {\bf 69}, 2863 (1992);
     Phys.\ Rev.\ B {\bf 48}, 10 345 (1993).}

\bibitem{white96}{S.\ R.\ White, 
     Phys.\ Rev.\ B.\ {\bf 53}, 52, (1996).}

\bibitem{morewhite}{S.~R. White, R. Noack, and D.~J. Scalapino,
    Phys.\ Rev.\ Lett.\ {\bf 73}, 886 (1994).}

\bibitem{poilblanc}{D. Poilblanc, H. Tsunetsugu, and T.~M. Rice,
    Phys.\ Rev.\ B {\bf 50}, 6511 (1994).}

\bibitem{dennijs}{M.\ P.\ M.\ den Nijs, 
     Phys.\ Rev.\ B {\bf 23}, 6111 (1981).}

\bibitem{luther}{A.\ Luther and I.\ Peschel, 
      Phys.\ Rev.\ B {\bf 12}, 3908 (1975).
      F.\ D.\ M.\ Haldane, 
      Phys.\ Rev.\ Lett.\ {\bf 45}, 1358 (1980).}

\bibitem{tsvelik}{A.\ M.\ Tsvelik, {\it Quantum Field Theory in Condensed 
      Matter Physics}, Cambridge University Press (1995).}

\bibitem{affleck}{I.\ Affleck, in {\it Fields, Strings and Critical Phenomena},
     edited by E.\ Br\'ezin and J.\ Zinn-Justin, North-Holland,
     Amsterdam (1990).}

\bibitem{hsiuhau}{H.\ H.\ Lin, L.\ Balents, and M.\ P.\ A.\ Fisher, 
      Phys. Rev. B {\bf 56}, 6569 (1997).}

\bibitem{orignac}{E. Orignac and T. Giamarchi, 
    Phys.\ Rev.\ B {\bf 56}, 7167 (1997).}

\bibitem{jose}{J.~V. Jos\'{e}, L.~P. Kadanoff, S. Kirkpatrick, and
    D.~R. Nelson, Phys, Rev. B {\bf 16} 1217, (1977).}
       
\bibitem{giaschulz}{T.\ Giamarchi and H.\ J.\ Schulz, 
      Phys. Rev. B {\bf 37}, 325, (1988).}


\bibitem{konik}{R.\ M.\ Konik and A.\ LeClair, 
       Nucl. Phys. B {\bf 479}, 619, (1996).}

\bibitem{boyanovsky}{D.\ Boyanovsky, 
       J.\ Phys.\ A {\bf 22}, 2601, (1989).}


\end{references}
\end{document}